%% file: main.tex
\documentclass[11pt,a4paper]{article}
\usepackage{times}
\usepackage{latexsym}
\usepackage{microtype}

\usepackage[final]{emnlp2021}

\usepackage[utf8]{inputenc} 
\usepackage[T1]{fontenc}    
\usepackage{hyperref}       
\usepackage{url}            
\usepackage{float}
\usepackage{booktabs} 
\usepackage{array,multirow,textcomp}
\usepackage{xcolor} 
\usepackage{graphicx}
\usepackage{CJKutf8}
\usepackage{comment}

\usepackage{multirow}
\usepackage{ragged2e}
\usepackage{amsmath}

\usepackage[ruled,vlined,linesnumbered]{algorithm2e}
\usepackage{amsfonts}       
\usepackage{nicefrac}       
\usepackage{microtype}      
\usepackage{enumitem}
\usepackage{xcolor,colortbl}

\newcommand{\bm}{\textbf}

\definecolor{lightgray}{rgb}{0.83, 0.83, 0.83}
\definecolor{darkgray}{rgb}{0.5, 0.5, 0.5}
\definecolor{mediumgray}{rgb}{0.66, 0.66, 0.66}

\newcommand{\testname}{ValNorm}

\title{Instructions for ACL-IJCNLP 2021 Proceedings}

\author{Autumn Toney-Wails \\
  Georgetown University  \\
  \texttt{autumn.toney@georgetown.edu} \\\And
  Aylin Caliskan \\
  University of Washington \\
  \texttt{aylin@uw.edu} \\}

\date{}

\begin{document}

\title{ValNorm Quantifies Semantics to Reveal Consistent Valence Biases Across Languages and Over Centuries}

\maketitle
\input{0_absract/abstract}
\input{1_introduction/introduction}

\input{2_relatedwork/relatedwork}

\input{3_data/data}

\input{4_methodology/methodology}

\input{5_experiments/experiments}

\input{6_results/results}

\input{7_discussion/discussion}

\input{8_conclusion/conclusion}

\bibliographystyle{ACM-Reference-Format}
\bibliography{references}



\input{appendix}

\end{document}

%% file: 0_absract/abstract.tex
\begin{abstract}

Word embeddings learn implicit biases from linguistic regularities captured by word co-occurrence statistics. By extending methods that quantify human-like biases in word embeddings, we introduce \testname{}, a novel intrinsic evaluation task and method to quantify the valence dimension of affect in human-rated word sets from social psychology. We apply \testname{} on static word embeddings from seven languages (Chinese, English, German, Polish, Portuguese, Spanish, and Turkish)  and from historical English text spanning 200 years. \testname{} achieves consistently high accuracy in quantifying the valence of non-discriminatory, non-social group word sets. Specifically, \testname{} achieves a Pearson correlation of $\rho=0.88$ for human judgment scores of valence for 399 words collected to establish pleasantness norms in English. In contrast, we measure gender stereotypes using the same set of word embeddings and find that social biases vary across languages. Our results indicate that valence associations of non-discriminatory, non-social group words represent widely-shared associations, in seven languages and over 200 years.

\end{abstract}

%% file: 1_introduction/introduction.tex
\section{Introduction}

New transparency-enhancing methods for static word embedding evaluation incorporate cross-disciplinary techniques that can quantify widely-accepted, intrinsic characteristics of words \cite{Bakarov, faruqui-etal-2016-problems, Schnabel2015, Hollenstein}. A promising approach for developing transparent intrinsic evaluation tasks is to evaluate word embeddings through the lens of \emph{cognitive lexical semantics}, which captures the social and psychological responses of humans to words and language \cite{Hollenstein, osgood1964semantic, osgood1975cross}. Such an approach could provide a representativeness evaluation method for word embeddings used in quantifying and studying biases. 

This paper presents the computational approach \testname{}, that accurately quantifies the valence dimension of biases and affective meaning in word embeddings, to analyze widely shared associations of non-social group words. Implicit biases, as well as the intrinsic pleasantness or goodness of things, namely valence, have been well researched with human subjects \cite{Greenwald1998MeasuringID, russell1983pancultural, russell}. Valence is one of the principal dimensions of affect and cognitive heuristics that shape attitudes and biases in humans \cite{harmon2013does}. Valence is described as the affective quality referring to the intrinsic attractiveness/goodness or averseness/badness of an event, object, or situation \cite{frijda1986emotions,  osgood1957measurement}. For word embeddings, we define valence bias as the semantic evaluation of pleasantness or unpleasantness that is associated with words (e.g., \textit{kindness} is associated with pleasantness and \textit{torture} is associated with unpleasantness).

Word embedding evaluation tasks are methods to measure the quality and accuracy of learned word vector representations from a given text corpus. The two main types of evaluation tasks are \textit{intrinsic evaluation}, which analyzes and interprets the semantic or syntactic characteristics of word embeddings (e.g., word similarity), and  \textit{extrinsic evaluation}, which measures how well word embeddings perform on downstream tasks (e.g., part-of-speech tagging, sentiment classification) \cite{Wang, Tsvetkov}. We focus on intrinsic evaluation, specifically the semantic quality of word embeddings that have been shown to learn human-like biases, such as gender and racial stereotypes \cite{caliskan}.

Our intrinsic evaluation task, \testname{}, accurately quantifies the valence dimension of biases and affective meaning in word embeddings. Simply, \testname{} provides a statistical measure of the pleasant/unpleasant connotation of a word. We validate \testname{}'s ability to quantify the semantic quality of words by implementing the task on word embeddings from seven different languages (Chinese, English, German, Polish, Portuguese, Spanish, and Turkish) and over a time span of 200 years (English only). Our results showcase that non-discriminatory non-social group biases introduced by \citet{bellezza1986words} are consistent across cultures and over time; people agree that loyalty is pleasant and that hatred is unpleasant. Additionally, we use the Word Embedding Association Test (WEAT) \cite{caliskan} to measure the difference between social group biases and non-discriminatory biases in seven languages.

We compare \testname{} to six widely used, traditional intrinsic evaluation tasks that measure how semantically similar two words are to each other and how words relate to each other. All intrinsic evaluation tasks, including \testname{}, measure the correlation of the computed scores to human-annotated scores. We implement all intrinsic evaluation tasks on seven word embedding sets (English only), which were trained using four different embedding algorithms and five different training text corpora (see Figure~\ref{fig:intrins results}) to ensure that the \testname{} results are not model or corpus specific. \testname{} achieves Pearson correlation coefficients ($\rho$) in the range $[0.82, 0.88]$ for the seven English word embedding sets, outperforming the six traditional intrinsic evaluation tasks we compare our results to.

We summarize our three main contributions: \textbf{1)} We quantify semantics, specifically the valence dimension of affect (pleasantness/unpleasantness) to study the valence norms of words and present a permutation test to measure the statistical significance of our valence quantification, \textbf{2)} we introduce \testname{}, a new intrinsic evaluation task that measures the semantic quality of word embeddings (validated on seven languages), and \textbf{3)} we establish widely-shared associations of valence across languages and over time.
Extended methodology, results, dataset details are in the appendices; the open source repository will be made public.

%% file: 2_relatedwork/relatedwork.tex
\section{Related Work}

Derived from the Implicit Association Test (IAT) in social psychology, Caliskan et al. defined the Word Embedding Association Test (WEAT) and the Word Embedding Factual Association Test (the single-category WEAT), to measure implicit biases in word embeddings \cite{caliskan}. The WEAT has two tests that measure non-social group (e.g., flowers) biases and seven tests that measure social group (e.g., gender, race) biases. The social group WEATs have been widely studied in the natural language processing (NLP) domain, as understanding social group biases is important for society. The single-category WEAT (SC-WEAT) measured gender bias in occupations and androgynous names, which highly correlate with gender statistics \cite{caliskan}.

SC-WEAT resembles a single-category association test in human cognition \cite{caliskan2020social, guo2021detecting, karpinski2006single}. SC-WEAT also shares similar properties with lexicon induction methods, which automatically extract semantic dictionaries from textual corpora without relying on large-scale annotated data for training machine learning models \cite{hatzivassiloglou1997predicting}. \citet{riloff2003learning, turney2003measuring} apply lexicon induction methods for sentiment, polarity, orientation, and subjectivity classification.

In prior work, the classification of valence is not evaluated in the context of measuring the quality of word embeddings or quantifying valence norms. 

\citet{lewis2020gender} investigate the distributional structure of natural language semantics in 25 different languages to determine the gender bias in each culture. While Lewis and Lupyan analyze bias across languages, they focus specifically on the social group of gender, and not on widely shared associations across languages. Garg et al. quantify gender and ethnic bias over 100 years to dynamically measure how biases evolve over time \cite{garg2018word}. Similarly, Garg et al. do not measure widely shared associations over time, they only measure social group biases.

Predicting affective ratings of words from word embeddings has proven to be a more complex task than computing word similarity, and is typically approached as a supervised machine learning problem \cite{li, teofili2019affect, wang2016}. Affect ratings of words computed from word embeddings can improve NLP tasks involving sentiment analysis and emotion detection \cite{Ungar2017PredictingEW, mohammad2016sentiment}, thus, designing an intrinsic evaluation task that estimates the valence association of a word is significant.

Traditional word embedding intrinsic evaluation tasks use word similarity, word analogy, or word categorization to measure linguistic properties captured by word embeddings \cite{Schnabel2015, Bakarov}. Word similarity and word analogy tasks use cosine similarity to measure semantic similarity of the vector representations of words in the evaluation task. Word similarity tasks compare the cosine similarity to a human-rated similarity score through Pearson or Spearman correlation \cite{pearson, spearman}; the correlation coefficient provides the accuracy metric for semantic similarity learned by word embeddings. Word analogy tasks output matching words based on vector arithmetic and accuracy is the metric of correct word selection \cite{Mikolov2013EfficientEO}. Since there is no standardized approach to evaluate word embeddings, we focus on the five most commonly used word similarity tasks WordSim (353 word pairs), RareWord (2,034 word pairs), MEN (3,000 word pairs), SimLex (999 word pairs), SimVerb (3,500 word pairs) \cite{finkelstein2001placing, luong2013better, bruni2014multimodal, hill2015simlex, gerz2016simverb, turney2010frequency}, and the word analogy task from \citet{Mikolov2013EfficientEO} which contains 8,869 semantic and 10,675 syntactic questions.

%% file: 3_data/data.tex
\section{Datasets}
\label{sec:dataset}
We use two main sources of data: 1) word embeddings and 2) human-annotated validation datasets.

\textbf{Word Embeddings:} 
We choose six widely-used, pre-trained word embedding sets in English, listed in Table~\ref{tab:word_embeddings}, to compare \testname{}'s performance on different algorithms (GloVe, fastText, word2vec) and training corpora (Common Crawl, Wikipedia, OpenSubtitles, Twitter, and Google News) \cite{pennington, Bojanowski, Mikolov2013EfficientEO, grave2018learning}. We include a seventh word embedding set, ConceptNet Numberbatch, since it is comprised of an ensemble of lexical data sources and is claimed to be less prejudiced in terms of ethnicity, religion, and gender \cite{speer2017conceptnet}. ConceptNet Numberbatch's results on social group and non-social group association tests provide a unique insight into valence norms for word embeddings, since the social group biases have been intentionally lowered.

\begin{table}[h]\small
  \begin{tabular}{lll} \\ \hline
     
     \textbf{Algorithm} & \textbf{Corpus} & \textbf{\# Tokens} \\ \hline
     
    \multirow{2}{*}{ \shortstack{fastText skipgram}} & Common Crawl  & 600B \\ 
    & OpenSubtitles 2018  &  22.2B\\ \hline
    
     \multirow{2}{*}{ \shortstack{fastText CBOW}} & Common Crawl  & \multirow{2}{*}{600+ B} \\
        & Wikipedia & \\ \hline
  
    \multirow{2}{*}{ \shortstack{GloVe}} & Common Crawl & 840B \\ 
    & Twitter & 27B \\ \hline
    Numberbatch & - & - \\ \hline
    word2vec & Google News & 100B \\\hline 
    \multicolumn{3}{p{22em}}{\footnotesize{300-dimensional embeddings (except the 200-dimensional GloVe Twitter)}}
    \end{tabular}
    
\captionof{table}{Details of word embeddings used in experiments; the corpus and token count for Numberbatch word embeddings are blank since they are formed using lexical information from ConceptNet, OpenSubtitles 2016, GloVe, and word2vec \cite{speer2017conceptnet}.
}
\label{tab:word_embeddings}
\end{table}

We use the 300-dimensional, pre-trained fastText word embeddings prepared for 157 languages for our seven languages of interest from five branches of language families that have different syntactic properties (Chinese, English, German, Polish, Portuguese, Spanish, and Turkish) \cite{grave2018learning}. 

For longitudinal valence analysis, we use historical word embeddings from \citet{hamilton2016diachronic} trained on English text between 1800 and 1990. Each word embedding set covers a 10-year period.

\textbf{Validation Datasets (English):} We choose three validation, human-annotated datasets of varying size for our experiments in English. All human-rated valence scores are reported as the mean. 

\citet{bellezza1986words} compiled a vocabulary list of 399 words to establish norms for pleasantness, imagery, and familiarity. College students rated words on pleasantness versus unpleasantness, which corresponds to cognitive representation of valence. The Affective Norms for English Words (ANEW) dataset is a widely used resource in sentiment analysis for NLP tasks. ANEW contains 1,034 vocabulary words and their corresponding valence, arousal, and dominance ratings. Psychology students were asked to rate words according to the Self-Assessment Manikin (SAM), on a scale of 1 (unhappy) to 9 (happy) \cite{bradley1999affective}. \citet{warriner} extended ANEW to 13,915 vocabulary words by adding words from more category norms (e.g., taboo words, occupations, and types of diseases). 1,827 Amazon Mechanical Turk workers rated words on the SAM scale of 1 to 9. \citet{warriner} note that valence scores were comparatively similar among responses.

The three human-annotated datasets have 381 common words in their respective vocabularies\footnote{The missing words are `affectionate', `anxiety', `capacity', `comparison', `constipation', `disappointment', `easter', `epilepsy', `hitler', `inconsiderate', `magnificent', `me', `nazi', `prosperity', `reformatory', `sentimental', `tuberculosis', `woman'.}. Using this subset of 381 words, we measure the Pearson correlation ($\rho$) of human valence scores across all three datasets to assess the inter-rater reliability. Our measurements result in $\rho \geq 0.97$ for all combinations of comparison. This high correlation indicates a strong inter-rater reliability for valence scores of words and signals widely shared associations, since each dataset was collected from a different year (1995, 1999, and 2013) with different groups of participants from various backgrounds.

\textbf{Validation Datasets (Cross-linguistic):}

ANEW has been adapted to many languages in order to interpret affective norms across cultures. We select five adaptations of ANEW: German, Polish, Portuguese, Spanish, and Turkish. We found these sets to be most complete (included majority of the ANEW vocabulary) and representative of various language structures (e.g., Turkish is a non-gendered language). We also include an affective norm Chinese dataset that contains a large overlapping vocabulary, but is not an ANEW adaptation.

The Polish, Portuguese, Spanish, and Turkish adaptations of ANEW use the original set of words, translated by experts to ensure accurate cross-linguistic results, and collected human-rated valence scores on the SAM 1 to 9 point scale for unhappy/happy according to the original ANEW study \cite{imbir2015affective, soares2012adaptation, redondo2007spanish, kapucu2018turkish}. The German adaptation of ANEW is an extension of the Berlin Affected Word List (BAWL)\cite{vo2009berlin} and was relabeled as Affective Norms for German Sentiment Terms (ANGST) \cite{schmidtke2014angst}. Valence scores were collected on a -3 (unhappy) to 3 (happy) point scale \cite{schmidtke2014angst}. The Chinese Valence-Arousal Words (CVAW) contains 5,512 words rated by four expert annotators who were trained on the Circumplex Model of Affect, which is one the foundational methodologies for affective meaning of words \cite{russell, yu-etal-2016-building}. The annotators assigned sentiment scores on a 1 (negative) to 9 (positive) point scale accordingly \cite{yu-etal-2016-building}. 

\begin{center}
\begin{tabular}{p{0.4cm}p{0.4cm}p{0.4cm}p{0.4cm}p{0.4cm}p{0.4cm}p{0.4cm}}
    \toprule
   & \textbf{EN} & \textbf{DE} & \textbf{PL} & \textbf{PT} & \textbf{ES} & \textbf{TR}  \\ 
  \hline
  \textbf{CN}  & 0.91 & 0.85 & 0.88 & 0.88 & 0.89 & 0.83 \\ \hline
  \textbf{EN}   & 1.0 & 0.95 & 0.94 & 0.95 & 0.95 & 0.87 \\ \hline
  \textbf{DE}  &  & 1.0 & 0.92 & 0.94 & 0.95 & 0.87 \\ \hline
  \textbf{PL} & & & 1.0 & 0.93 & 0.93 & 0.85\\ \hline
  \textbf{PT} & & & & 1.0 & 0.96 & 0.88\\ \hline
  \textbf{ES} & & & &  & 1.0 & 0.88\\
  \bottomrule
    \end{tabular}
\captionof{table}{Pearson correlation coefficient ($\rho$) of human scores in the validation sets across languages.}
\label{tab:lang rater corr} 
\end{center}

We identify the words in common across the seven cross-linguistic datasets, and we check the variance in human-annotated valence scores for this subset of 143 words\footnote{This word set is mainly limited by the Chinese dataset. See Table \ref{tab: wefat_foreign} for the number of words contained in each language's dataset.}. The top five words with the \textbf{least} amount of variance in valence are \textit{terrific} (\bm{$\sigma^2$} $= 3.6 \times 10^{-4}$), \textit{loyal} (\bm{$\sigma^2$} $= 4.7 \times 10^{-4}$), \textit{humor} (\bm{$\sigma^2$} $= 6.9 \times 10^{-4}$), \textit{hatred} (\bm{$\sigma^2$} $= 7.4 \times 10^{-4}$), and \textit{depression} (\bm{$\sigma^2$} $= 9.2 \times 10^{-4}$). The top five words with the \textbf{most} amount of variance in valence are \textit{execution} (\bm{$\sigma^2$} $= 4.9 \times 10^{-2}$), \textit{party} (\bm{$\sigma^2$} $= 4.1 \times 10^{-2}$), \textit{vomit} (\bm{$\sigma^2$} $= 3.4 \times 10^{-2}$), \textit{malaria} (\bm{$\sigma^2$} $= 2.7 \times 10^{-2}$), and \textit{torture} (\bm{$\sigma^2$} $= 2.6 \times 10^{-2}$). The overall variance of valence for all words are low.

%% file: 4_methodology/methodology.tex
\section{Methods}
\label{approach}
To measure social group biases and valence norms, we use the Word Embedding Association Test (WEAT) and the single-category Word Embedding Association Test (SC-WEAT).

\subsection{Association Tests: WEAT \& SC-WEAT}
The WEAT and SC-WEAT compute an effect-size statistic (Cohen's $d$) \cite{cohen2013statistical} measuring the association of a given set of target words or a single vocabulary word between two given attribute sets in a semantic vector space composed of word embeddings. The WEAT measures the differential association between two sets of target words and two sets of polar attribute sets, and the SC-WEAT measures the association of a single word to the two sets of polar attributes. Stimuli representing target social groups and polar attributes used in the WEAT are borrowed from the IATs designed by experts in social psychology. Table~\ref{tab:equations} provides the equations to compute the effect sizes for WEAT and SC-WEAT and their respective $p$-values; the $p$-values represent the significance of the effect sizes. $|d| \geq 0.80$ represents a biased association with high effect size \cite{cohen2013statistical}, with a one sided $p$-value $\leq 0.05$ or $p$-value $\geq 0.95$ representing a statistically significant effect size. 
    \begin{table*}[t]
    \centering
    \resizebox{\textwidth}{!}{%
     \begin{tabular}{ l l  l } \hline

\textbf{Test} & \textbf{Effect Size}  ($es$ measured in Cohen's $d$) & \bm{$p$}\textbf{-value}  \\ \hline \\

\textbf{WEAT} 
& $es(X, Y, A, B) =  \frac{\textrm{mean}_{x \in X} s(x, A, B) - \textrm{mean}_{y \in Y}s(y, A, B)}{\textrm{std-dev}_{w \in X \cup Y}s(w, A, B)}$ 
& $\Pr_{i}[s(X_{i}, Y_{i}, A, B) > s(X, Y, A, B)]$ \\ \\ \hline 
\\

    \textbf{SC-WEAT} & 
$es(\vec{w}, A, B) = \frac{\textrm{mean}_{a \in A} \textrm{cos}(\vec{w}, \vec{a}) - \textrm{mean}_{b \in B} \textrm{cos}(\vec{w}, \vec{b})}{\textrm{std-dev}_{x \in A \cup B}\textrm{cos}(\vec{w}, \vec{x})}$ 
& $\Pr_{i}[s(\vec{w}, A_{i}, B_{i}) > s(\vec{w}, A, B)]$ \\ 
 \\ \hline

  \multicolumn{3}{p {14cm}}{ \small{\textbf{Association:}   $ s(w, A, B) = \textrm{mean}_{a \in A} \textrm{cos}(\vec{w}, \vec{a}) - \textrm{mean}_{b \in B} \textrm{cos}(\vec{w}, \vec{b})$}} \\

 \multicolumn{3}{p {14cm}}{ \small{\textbf{Cosine similarity:} $\textrm{cos}(\vec{a}, \vec{b})$ denotes the cosine of the angle between the vectors $\vec{a}$ and $\vec{b}$.}} \\

\multicolumn{3}{p {17cm}}{\small{\textbf{Target words:} $X = [\vec{x_{1}}, \vec{x_{2}},\dots,\vec{x_{m}}]$ and $Y = [\vec{y_{1}}, \vec{y_{2}},\dots,\vec{y_{m}}]$ are the two equal-sized ($m$) sets of target stimuli.  }} \\ 

\multicolumn{3}{p {14cm}}{\small{\textbf{Attribute words:} $A = [\vec{a_{1}}, \vec{a_{2}},\dots,\vec{a_{n}}]$ and $B = [\vec{b_{1}}, \vec{b_{2}},\dots,\vec{b_{n}}]$ are the two equal-sized ($n$) sets of attributes.}} \\

         \multicolumn{3}{p {14cm}}{\small{\textbf{SC-WEAT}: $\vec{w}$ is the single target stimulus in SC-WEAT.}} \\

  \multicolumn{3}{p {16cm}}{\small{\textbf {Permutation test:} $(X_i, Y_i)_i$ and $(A_i, B_i)_i$ denote all the partitions of $X \cup Y$ and $A \cup B$ into two sets of equal size. Random permutations of these sets represent the null hypothesis  as if the biased associations did not exist so that we can perform a statistical significance test by measuring the unlikelihood of the null hypothesis, given the effect size of WEAT or SC-WEAT. }} \\

    \end{tabular}
    }
    \caption{WEAT and SC-WEAT effect size equations and their corresponding statistical significance in $p$-values.}
    \label{tab:equations}
    \end{table*}
    
We use the WEAT and SC-WEAT to quantify biases and measure statistical regularities in text corpora. We extend the SC-WEAT to precisely measure valence using pleasant/unpleasant evaluative attribute sets provided by \citet{Greenwald1998MeasuringID} (as opposed to male/female from \citet{caliskan}). We design our intrinsic evaluation task, \testname{}, around this valence quantification method. We extend all methods (WEAT, SC-WEAT, \testname{}) to six non-English languages using native speaker translations of the word sets. In all of our experiments we use the defined sets of stimuli from \cite{caliskan} to ensure that our experiments provide accurate results. Following WEAT, each word set contains at least 8 words to satisfy concept representation significance. Accordingly, the limitations of not following WEAT's methodological robustness rules, which are analyzed by \citet{ethayarajh2019understanding}, are mitigated.

\subsection{Statistical Significance of Valence Quantification} 

Caliskan et al. do not present a $p$-value for the SC-WEAT effect size\footnote{Caliskan et al. measure the $p$-value of the correlation between the SC-WEAT computed gender association scores and their corresponding ground truth values obtained from annual U.S. Census and Labour Bureau statistics.}. Thus, we define the one-sided $p$-value of SC-WEAT where $\{(A_{i}, B_{i})\}_{i}$ represents the set of all possible partitions of the attributes $A \cup B$ of equal size to represent the null hypothesis. The null hypothesis is, for a given stimulus $\vec{w}$, computing the SC-WEAT effect size using a random partition of the attribute words $\{(A_{i}, B_{i})\}_{i}$ represents the empirical distribution of effect sizes in case there were no biased associations between the stimulus and the attribute sets. Accordingly, the permutation test measures the unlikelihood of the null hypothesis for SC-WEAT.

\subsection{\testname{}: An Intrinsic Evaluation Task}
Our intrinsic evaluation task uses the SC-WEAT with pleasant and unpleasant attribute sets to represent the valence dimension of affect\footnote{ \url{https://github.com/autumntoney/ValNorm}}. \testname{}'s output is the Pearson's correlation value when comparing the computed valence scores to the human-rated valence scores from a ground truth validation dataset. We define the \testname{} task as:\\
\noindent 1. \textbf{Assign} the \textit{word} column from the validation dataset to $W$, the set of target word vectors.\\
\noindent 2. \textbf{Assign} the pleasant attribute words to $A$, the first attribute set, and assign the unpleasant attribute words to $B$ the second attribute set.\\
\noindent    3. \textbf{Compute} SC-WEAT effect size and $p$-value for each $\vec{w} \in W$ using the given word embedding set.\\
\noindent    4. \textbf{Compare} SC-WEAT effect sizes to the human-rated valence scores using Pearson's correlation to measure the semantic quality of word embeddings.

For non-English languages we include a preliminary step to \testname{}, where we translate the pleasant/unpleasant attribute word sets from English to the given language and verify the translations with native speakers.

\subsection{Discovering Widely-Accepted Non-Social Group Associations}
We investigate the existence of widely-accepted non-social group biases by implementing the \textit{flowers-insects-attitude}, \textit{instruments-weapons-attitude}, and \textit{gender-science} WEATs defined by \citet{caliskan} on word embeddings from our seven languages of interest\footnote{Cross-linguistic `flowers-insects', `instruments-weapons', and `gender-science' WEATs have been replicated using their corresponding attribute word sets from IATs on Project Implicit's \cite{nosek2002harvesting, nosek2009national} webpages in the seven languages we analyzed (Chinese, English, German, Polish, Portuguese, Spanish, and Turkish).}. Both non-social group attitude tests are introduced as `universally accepted stereotypes' in the original paper that presents the IAT \cite{Greenwald1998MeasuringID}. Thus, these are baseline biases that we expect to observe with high effect size in any representative word embeddings. We use the \textit{gender-science} WEAT results, which measures social group biases, to compare with our non-social group bias tests' results. In this way, we can identify if social and non-social group biases have consistent results across languages to infer if our non-social group bias results indicate universality.

Additionally, we implement \testname{} on the six non-English word embedding sets and historical word embeddings from 1800--1990.

%% file: 5_experiments/experiments.tex
\section{Experiments}
\label{sec:experiments}

We conduct three main experiments to 1) quantify the valence statistics of words in text corpora, 2) evaluate our intrinsic evaluation task, \testname{} and 3) investigate widely-shared non-social group valence associations across languages and over time.

\subsection{Quantifying Valence Using SC-WEAT} We use the SC-WEAT to quantify valence norms of words by measuring a single word's relative association to pleasant versus unpleasant attribute sets. We use the same word sets of 25 pleasant and 25 unpleasant words used in \citet{caliskan} flowers-insects-attitude bias test. These attribute word sets were designated by experts in social psychology to have consistent valence scores among humans \cite{Greenwald1998MeasuringID}. We run the SC-WEAT on the seven sets of word embeddings listed in Section~\ref{sec:dataset}, and we evaluate each word embedding set using valence lexica.

\subsection{Evaluating \testname{}}
We run \testname{} on the seven English word embedding sets, using Bellezza's Lexicon, ANEW, and Warriner's Lexicon as the target word set respectively. 
We measure the correlation of the \testname{} scores to the corresponding set of human-rated scores. We compare ValNorm's results to the results from six traditional intrinsic evaluation tasks on the seven English word embedding sets. This evaluation compares six traditional evaluation tasks to three implementations of \testname{} across seven sets of word embeddings, trained using four different algorithms and five different text corpora.

To investigate the significance of training corpus size for word embeddings, we sample 5 bin sizes (50\%, 10\%, 1\%, 0.1\%, and 0.001\%) of the OpenSubtitles 2018 corpus and train word embeddings according to \citet{Paridon}'s method to generate subs2vec (fastText skipgram 300-dimensional word embeddings). We choose the OpenSubtitles corpus for this experiment since it reflects human communication behavior more closely than a structured written corpus, such as Wikipedia or news articles, making it a more appropriate corpus for capturing semantic content \cite{Paridon}.

There are 89,135,344 lines in the cleaned and deduplicated OpenSubtitles corpus text file, which we round to 89,000,000 to make our sample size bins neat. For each bin size we randomly sample, without replacement, the designated number of lines in the text corpus file. We generate word embeddings for each sample size and run the five word similarity intrinsic evaluation tasks and the \testname{} evaluation task to analyze the significance of corpus size on word embedding quality.

\begin{figure*}[ht!]
    \centering
      \includegraphics[width = \textwidth]{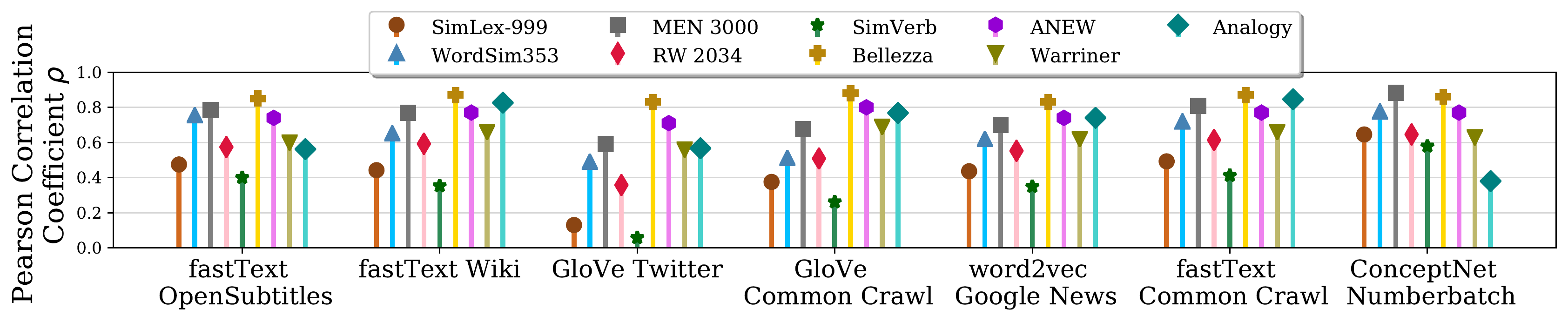}
          
    \caption{Comparison of nine intrinsic evaluation tasks on seven widely used word embeddings shows that \testname{} achieves the highest correlation with human-rated scores, outperforming other intrinsic evaluation metrics.} 
    \label{fig:intrins results}
\end{figure*}

\subsection{Analyzing Widely Shared Associations} 
We use the WEAT to quantify valence associations of non-social groups (flowers, insects, instruments, and weapons) and to quantify social group (male/female) associations to science and arts. We hypothesize that valence biases will remain consistent across word embeddings, and that social group biases will change. Gender bias scores in word embeddings may vary depending on culture and language structure (e.g., Turkish pronouns are gender-neutral). We compare the result differences from the valence association tests and the gender association test on seven different sets of English word embeddings (see Table~\ref{tab:word_embeddings}) and on word embeddings from six other languages (Chinese, German, Polish, Portuguese, Spanish, and Turkish). We were unable to run these WEATs on the historical word embeddings, as their vocabularies did not contain most of the target and attribute words. 

We implement \testname{} across the six non-English languages, using Bellezza's Lexicon as the target set, since all languages (except for Chinese) had at least 97\% of the words in their ground-truth dataset (see Table~\ref{tab: wefat_foreign}). We also evaluate the stability of valence norms over 200 years by implementing \testname{} on historical embeddings. If valence norms are independent of time, culture, and language, they will be consistent over 200 years and across languages, making them an appropriate metric for evaluating word embeddings.
\vspace{-3mm}

%% file: 6_results/results.tex
\section{Results}
\label{sec:results}

\textbf{Quantifying valence norms.} We implement the SC-WEAT using valence evaluative attributes and target word sets, that are hypothesized to represent valence norms, from Bellezza's Lexicon, ANEW, and Warriner's Lexicon. Our initial experiments signalled widely shared associations of valence scores with $\rho \in [0.82, 0.88]$ for all seven English word embeddings using Bellezza's Lexicon (vocabulary size of 399\footnote{The dataset section includes the details for words that are not included in cross-linguistic experiments.}) as the target word set. The corresponding $p$-values have a Spearman's correlation coefficient greater than $\rho \ge 0.99$ to the effect sizes, indicating statistically significant results.

\textbf{\testname{} performance.}
Figure~\ref{fig:intrins results} compares the performance of \testname{} using three valence lexica to five word similarity tasks and one analogy task. \testname{} using Bellezza's Lexicon overperforms all other intrinsic evaluation tasks on word embeddings trained on five corpora via four algorithms.

\textbf{Widely shared associations.} We compute the variance ($\sigma^2$) of the effect sizes for the flowers-insects-attitude, instruments-weapons-attitude, and gender-science WEAT bias tests across all seven language word embeddings. In Table \ref{tab:weat_variance}, as expected based on findings in social psychology, flowers-insects-attitudes and instruments-weapons have the most consistent valence associations, with $0.13$ and $0.09$ variance scores respectively.

\begin{figure*}[ht!]
    \centering
    \includegraphics[width = 0.98\textwidth]{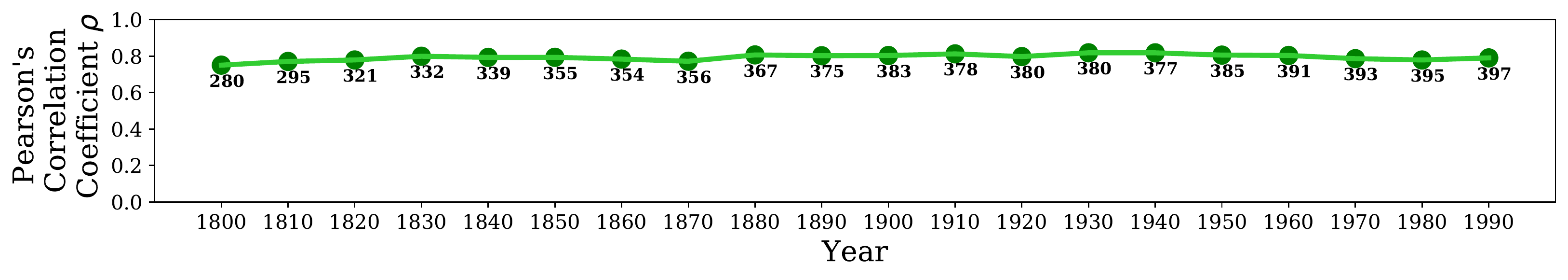}
    \vspace{-3mm}
    \caption{$\rho$ for \testname{} using Bellezza's Lexicon and HistWords historical word embeddings \cite{hamilton2016diachronic}. Points are labeled with the number of target words present in their vocabulary. The low variance ($\sigma^{2} < 10^{-3}$) of results validate our hypothesis that valence norms are consistent over two centuries. }
    \label{fig:timeline}

\end{figure*}

\begin{figure}[ht!]
\centering
        \includegraphics[width = 0.47\textwidth]{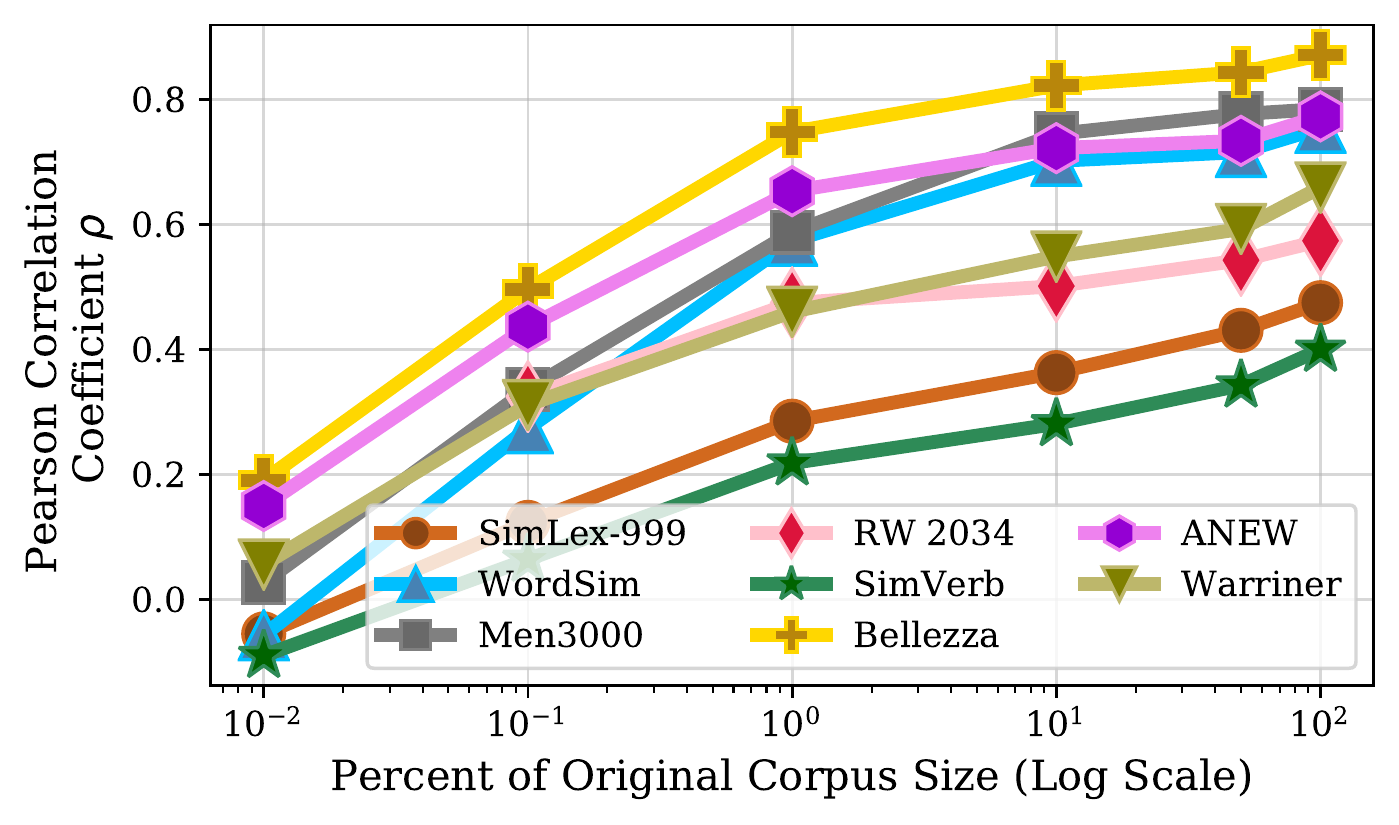}
            \vspace{-2mm}
        \captionof{figure}{Performance of word embeddings trained on the OpenSubtitles2018 corpus of different sizes.} 

        \label{fig:bin_eval}
\end{figure}

\begin{table}[h!]
\centering
    \vspace{-3mm}

\begin{tabular}{  l  r  } \\ \toprule
        \textbf{Bias Type} & \bm{$\sigma^2$}  \\ \midrule
   \hspace{-1mm}     flowers-insects & 0.13  \\ 
    \hspace{-1mm}      instruments-weapons & 0.09 \\ 
    \hspace{-1mm}      gender-science & 0.45 \\  \bottomrule
        \end{tabular}

          \caption{Variance of WEAT scores across 7 languages. The variances of the scores for widely accepted biases are low, whereas culture-specific social group bias scores measuring gender-science associations have high variance across languages. }
  \label{tab:weat_variance}

  \end{table}

Table \ref{tab: wefat_foreign} reports Pearson correlation coefficients using \testname{}, compared to the corresponding validation dataset for all seven languages, providing insight into consistent valence norms across cultures. Figure \ref{fig:timeline} shows the stability of valence norms over 200 years, with low variance in scores ($\sigma^{2} < 10^{-3}$), reporting the Pearson correlation coefficients for the valence association scores compared to the corresponding human-rated valence scores from Bellezza's Lexicon (compiled in 1986). Each point on the graph is labeled with the number of vocabulary words from Bellezza's lexicon that was present in the embedding's vocabulary; slight fluctuations in correlation scores may be dependent on the changes in words that were tested.

Figure \ref{fig:bin_eval} presents the results for the training corpus size experiment. For all intrinsic evaluation tasks the correlation score increases minimally from 50\% to 100\% and from 10\% to 50\%; \testname{} using Bellezza's Lexicon has a 0.01 and 0.03 increase respectively.

    \begin{table}[h!]

    \centering
    \begin{tabular}{ l p{10mm} p{10mm} } \toprule
        \textbf{Language} & \textbf{$N$} & \bm{$\rho$}  \\ 
        \midrule
        Chinese & 269 & 0.85 \\
        English & 381 & 0.87 \\
        EU Portuguese & 381 & 0.85  \\ 
        German & 370 & 0.80 \\ 
        Polish & 375 & 0.79 \\ 
        Spanish & 381 & 0.83 \\
        Turkish & 379 & 0.73 \\
        \bottomrule
            \end{tabular}
                
    \vspace{-2mm}

        \caption{Correlation coefficients ($\rho$) and number of target words present ($N$) in \testname{} for each language.}
    \label{tab: wefat_foreign}

   \end{table}

%% file: 7_discussion/discussion.tex
\section{Discussion} 
\label{sec:discussion}

In our three experiments we find evidence that word embeddings capture valence norms using \testname{} and WEAT to measure widely shared associations. These experiments show that valence norms relate to widely-shared associations, as opposed to culture specific associations, and can be used as a measurement of embedding quality across languages.

\textbf{\testname{} as a new intrinsic evaluation task.} Figure \ref{fig:intrins results} compares our three implementations of \testname{} to six traditional intrinsic evaluation tasks, with Bellezza's Lexicon performing the highest, most likely because it is designed specifically to measure valence norms and it is smaller than the other valence lexica. \testname{} computes an effect size rather than just the cosine similarity, the metric for word similarity and word analogy tasks. Notably, \testname{} using Bellezza's Lexicon (399 valence tasks) outperforms WordSim (353 similarity tasks). Using ANEW (1,035 valence tasks) and Warriner's Lexicon (13,915 valence tasks), \testname{} consistently outperforms SimLex (999 similarity tasks), RW (2,034 similarity tasks), and SimVerb (3,5000 similarity tasks). These results suggest that \testname{} measures valence accurately and consistently, regardless of the task size, whereas results of all other intrinsic evaluation tasks have high variance and lower accuracy. \testname{}'s performance supports our hypothesis that valence norms are captured by word co-occurrence statistics and that we can precisely quantify valence in word embeddings.


\textbf{Widely Shared Valence Associations.} Measuring \testname{} using Bellezza's Lexicon on Conceptnet Numberbatch word embeddings achieves $\rho = 0.86$. This high $\rho$ value highlights that, even when social group biases are reduced in word embeddings, valence norms remain and are independent of social group biases. The low variance of flowers-insects-attitude and instruments-weapons-attitude experiments signal widely-accepted associations for the non-social groups of flowers, insects, instruments, and weapons. Producing the highest variance of 0.45 across all languages, the gender-science experiment signals a culture and language specific association for gender social groups. 

Our corpus size experiment results using \testname{} follow the same trend-line as the other intrinsic evaluation tasks. This result signals widely-shared associations, since the co-occurrence statistics of the word embeddings preserve valence and word similarity scores comparably. When quantifying bias in embeddings, \testname{} can identify if the training corpus is of a sufficient size for representative and statistically significant bias analysis. 

Implementing \testname{} on seven different languages from five different language families, we find that valence norms are widely-shared across cultures. However, social group WEAT associations are not widely-shared; these results align with IAT findings from 34 countries \cite{nosek2009national}. Applying WEAT in seven languages, that belong to five branches of varying language families, shows that word embeddings capture grammatical gender along with gender bias. For example, when applying the gender-science WEAT in Polish by using the IAT words on Poland's Project Implicit, the resulting effect size signals stereotype-incongruent associations. Further analysis of this anomaly revealed that most of the words representing science in the Polish IAT have nouns with feminine grammatical gender. However, when the grammatical gender direction is isolated and removed from the word embeddings while performing WEAT, the results move to the stereotype-congruent direction reported via IATs on the Project Implicit site \cite{nosek2009national}. These findings suggest that structural properties of languages should be taken into account when performing bias measurements that might be somehow related to some syntactic property in a language. This analysis is left to future work since it does not directly affect valence norm measurements in language.

\testname{} quantifies stable valence norms over time with $\rho \in [0.75, 0.82]$ using historical word embeddings and Bellezza's Lexicon. While semantics are certainly evolving \cite{hamilton2016diachronic}, there are non-social group words that maintain their intrinsic characteristics at least for 200 years, as Bellezza et al. suggested, and furthermore, these words are consistent across languages.



%% file: 8_conclusion/conclusion.tex
\vspace{-1mm}
\section{Conclusion}
\label{sec:conclusion}
    \vspace{-2mm}

Valence norms reflect widely-shared associations across languages and time, offering a distinction between non-social group biases and social group biases (gender, race, etc.). These valence associations are captured in word embeddings trained on historical text corpora and from various languages. We document widely-shared non-social group associations as well as culture-specific associations via word embeddings. While the social group biases we measure vary, we find that non-social group valence norms are widely-shared across languages and cultures and stable over 200 years.

We present \testname{} as a new intrinsic evaluation task which measures the quality of word embeddings by quantifying the preservation of valence norms in a word embedding set. \testname{}, which has three implementations with increasing vocabulary sizes, outperforms traditional intrinsic evaluation tasks and provides a more sensitive evaluation metric based on effect size, as opposed to the cosine similarity metric of other evaluation tasks. Computationally quantifying valence of words produce a high correlation to human-rated valence scores, indicating that word embeddings can measure semantics, particularly valence, with high accuracy. The results of valence norms as statistical regularities in text corpora provides another layer of transparency into what word embeddings are learning during their training process.

    \vspace{-1mm}

\section{Ethical Considerations}
    \vspace{-2mm}

This work uses expert research in social psychology and computer and information science, specifically the Implicit Association Test (IAT) and the Word Embedding Association Test (WEAT), and applies it to the NLP domain in order to discover widely shared associations of non-discriminatory non-social group words \cite{Greenwald1998MeasuringID, caliskan}. Prior NLP applications of the WEAT focus mainly on social group biases, since studying potentially harmful features of machine learning and artificial intelligence (AI) are important for fair and ethical implementations of AI. Our application investigates valence (pleasant/unpleasant) associations that quantify attitudes, which can be used to analyze sentiment classification or for a more specific use case of detecting targeted language (information operations/hate speech). By establishing a method to measure valence norms, we establish an AI tool that can identify if biases in a text corpus align with widely accepted valence associations or if the language in the corpus expresses shifted biases.

While our work does not focus on social group biases and attitudes, valence association can be used as an indicator of how a social group is represented in text---is the group associated with pleasantness or unpleasantness? Social group biases are not consistent across cultures and over time, making this valence bias test useful in detecting derogatory or targeted attitudes towards social groups. It is also notable that we share valence associations regardless of language and culture; everyone agrees that kindness is pleasant and that vomit is unpleasant. This distinction between discriminatory biases (against social groups) and non-discriminatory biases (against non-social groups) creates a distinction in analyzing biases and stereotypes in languages. It may be acceptable if language expresses dislike of cancer, but harmful information may propagate to downstream applications if language expresses a negative attitude towards a specific race, gender, or any social group. 

\section{Acknowledgements}
We thank Osman Caliskan and Wei Guo for providing German and Chinese translations, respectively.

%% file: appendix.tex
\appendix
\begin{appendix}

\newpage

\section*{Appendix}

\begin{CJK*}{UTF8}{gbsn}
\begin{center}
\begin{table*}
\begin{tabular}{ |r l|p{0.7\textwidth}| } 
 \hline
  \multicolumn{2}{|c|}{\textbf{Category}}  &\textbf{Chinese Stimuli}  \\ 
 \hline 
 
 flowers & (target) & 三叶草, 兰花, 玫瑰, 水仙花, 紫丁香, 郁金香, 雏菊, 百合, 紫色, 木兰  \\  \hline
 insects & (target) & 蚂蚁, 跳蚤, 蜘蛛, 臭虫, 飞, 狼蛛, 蜜蜂, 蟑螂, 蚊子, 大黄蜂  \\  \hline
 instruments & (target) & 风笛, 大提琴, 吉他, 琵琶, 长号, 班卓琴, 单簧管, 口琴, 曼陀林, 喇叭, 巴松管, 鼓, 竖琴, 双簧管, 大号, 钟, 小提琴, 大键琴, 钢琴, 中提琴, 邦戈, 长笛, 喇叭, 萨克斯风, 小提琴\\  \hline
 weapons & (target) & 箭头, 俱乐部, 枪, 导弹, 矛, 斧头, 匕首, 鱼叉, 手枪, 剑, 刀, 炸药, 斧头, 步枪, 罐, 炸弹, 火器, 刀子, 滑膛枪, 催泪瓦斯, 大炮, 手榴弹, 锤, 弹弓, 鞭子 \\  \hline
 pleasant & (attributes) & 抚摸, 自由, 健康, 爱, 和平, 欢呼, 朋友, 天堂, 忠诚, 乐趣, 钻石, 温和, 诚实, 幸运, 彩虹, 文凭 \\  \hline
unpleasant & (attributes) & 滥用 , 崩溃 , 污秽 , 谋杀 , 疾病 , 事故 , 死亡 , 悲痛 , 毒 , 臭 , 突击 , 灾害 , 仇恨 , 污染 , 悲剧 , 离婚 , 监狱 , 贫穷 , 丑陋 , 癌症 , 杀 ,  烂 , 呕吐 , 痛苦 , 监狱\\  \hline
\end{tabular}
\caption{Chinese Stimuli (Word List)}
\label{tab:chinese}
\end{table*}
\end{center}
\end{CJK*}

\begin{center}
\begin{table*}
\begin{tabular}{ |r l|p{0.7\textwidth}| } 

 \hline
  \multicolumn{2}{|c|}{\textbf{Category}}  &\textbf{English Stimuli Collected from \citet{caliskan}}  \\ 
 \hline 
 flowers &(target) & clover, orchid, rose, daffodil, lilac, tulip, daisy, lily, violet, magnolia   \\ \hline
 insects & (target) & ant, flea, spider, bedbug, fly, tarantula, bee, cockroach, mosquito, hornet  \\ \hline
 instruments  &(target) & bagpipe,cello, guitar, lute, trombone, banjo, clarinet, harmonica, mandolin, trumpet, bassoon, drum, harp, oboe, tuba, bell, fiddle, harpsichord, piano, viola, bongo, flute, horn, saxophone, violin\\ \hline
 weapons &  (target) & arrow, club, gun, missile, spear, axe, dagger, harpoon, pistol, sword, blade, dynamite, hatchet, rifle, tank, bomb, firearm, knife, shotgun, teargas, cannon, grenade, mace, slingshot, whip \\ \hline
 pleasant &  (attributes) & caress, freedom, health, love, peace, cheer, friend, heaven, loyal, pleasure, diamond, gentle, honest, lucky, rainbow, diploma, gift, honor, miracle, sunrise, family, happy, laughter, paradise, vacation\\ \hline
unpleasant &  (attributes) & abuse , crash , filth , murder , sickness , accident , death, grief, poison, stink, assault, disaster, hatred, pollute, tragedy, divorce, jail, poverty, ugly, cancer, kill, rotten, vomit, agony, prison\\ 
 \hline
\end{tabular}
\caption{English Stimuli (Word List)}

\label{tab:english}
\end{table*}

\end{center}

\begin{center}
\begin{table*}
\begin{tabular}{ |r l|p{0.7\textwidth}| } 
 \hline
  \multicolumn{2}{|c|}{\textbf{Category}}  &\textbf{German Stimuli}  \\ 
 \hline 
 flowers&  (target) & Klee, Orchidee, Rose, Narzisse, Flieder, Tulpe, Gänseblümchen, Lilie, Veilchen, Magnolie  \\  \hline
 insects & (target) & Ameise, Floh, Spinne, Wanze, Fliege, Tarantel, Biene, Kakerlake, Mücke, Hornisse  \\  \hline
 instruments &  (target) & Dudelsack, Cello, Gitarre, Laute, Posaune, Banjo, Klarinette, Mundharmonika, Mandoline, Trompete, Fagott, Trommel, Harfe, Oboe, Tuba, Glocke, Geige, Cembalo, Klavier, Bratsche, Bongo, Flöte, Horn, Saxophon,  Violine\\  \hline
 weapons &  (target) & Pfeil, Keule, Waffe, Rakete, Speer, Axt, Dolch, Harpune, Pistole, Schwert, Klinge, Dynamit, Beil, Gewehr, Panzer, Bombe, Schusswaffe, Messer, Schrotflinte, Tränengas, Kanone, Granate, Streitkolben, Schleuder,  Peitsche\\  \hline
 pleasant &  (attributes) & Liebkosung, Freiheit, Gesundheit, Liebe, Frieden, Jubel, Freund, Himmel, Treue, Vergnügen, Diamant, sanft, ehrlich, glücklich, Regenbogen, Diplom, Geschenk, Ehre, Wunder, Sonnenaufgang, Familie, glücklich, Lachen, Paradies, Urlaub\\  \hline
unpleasant &  (attributes) & Missbrauch, Absturz, Schmutz, Mord, Krankheit, Unfall, Tod, Trauer, Gift, Gestank, Angriff, Katastrophe, Hass, Umweltverschmutzung, Tragödie, Scheidung, Gefängnis, Armut, hässlich, Krebs, töten, faul, Erbrechen, Qual, das Gefängnis\\  
 \hline
\end{tabular}
\caption{German Stimuli (Word List)}
\label{tab:german}
\end{table*}
\end{center}


\begin{center}
\begin{table*}
\begin{tabular}{ |r l|p{0.7\textwidth}| } 
 \hline
  \multicolumn{2}{|c|}{\textbf{Category}}&\textbf{Polish Stimuli}  \\ 
 \hline 
 flowers&  (target) & koniczyna, orchidea, róża, narcyz, liliowy, tulipan, stokrotka, lilia, fiołek, magnolia   \\  \hline
 insects & (target) & mrówka, pchła, pająk, pluskwa, latać, tarantula, pszczoła, karaluch, komar, szerszeń  \\  \hline
 instruments &  (target) & dudy,wiolonczela, gitara, flet, lutnia, puzon, banjo, klarnet, harmonijka, mandolina, trąbka, fagot, bęben, harfa, obój, tuba, dzwon, skrzypce, klawesyn, fortepian, altówka, bongo, róg, saksofon, skrzypce\\ \hline
 weapons &  (target) & strzałka, buława, strzelba, pocisk, włócznia, topór, sztylet, harpun, pistolet, miecz, nóż, dynamit, toporek, karabin, czołg, bomba, broń palna, ostrze, flinta, gaz łzawiący, armata, granat, buzdygan, proca, bat \\ \hline
 pleasant &  (attributes) & pieszczota, swoboda, zdrowie, miłość, dyplom, pokój, przyjemność, dopingować, przyjaciel, niebiosa, wierny, diament, delikatny, uczciwy, fartowny, tęcza, podarunek, honor, cud, rodzina, szczęśliwy, śmiech, raj, wakacje, świt\\ \hline
unpleasant &  (attributes) & nadużycie, wypadek, brud, zabójstwo, choroba, awaria, śmierć, smutek, trucizna, smród,atak,  katastrofa, nienawiść, zanieczyszczać, tragedia, rozwód, więzienie, bieda, brzydki, rak, zgniły, wymiociny, agonia, areszt, zło\\ 
 \hline
\end{tabular}
\caption{Polish Stimuli (Word List)}
\label{tab:polish}
\end{table*}

\end{center}


\begin{center}
\begin{table*}
\begin{tabular}{ |r l|p{0.7\textwidth}| } 
 \hline
  \multicolumn{2}{|c|}{\textbf{Category}}&\textbf{Portuguese Stimuli}  \\ 
 \hline 
 flowers&  (target) & trevo, orquídea, rosas, narciso, lilás, tulipa, margarida, lírio, tolet, magnólia  \\  \hline
 insects & (target) & formiga, pulga, aranha, percevejo, mosca, tarântula, abelha, barata, mosquito, vespa  \\  \hline
 instruments &  (target) & gaita de foles, violoncelo, violão, alaúde, trombone, banjo, clarinete, harmônica, bandolim, superada, fagote, tambor, harpa, oboé, tuba, sino, rabeca, cravo, piano, viola, bongo, flauta, chifre, saxofone, violino\\ \hline
 weapons &  (target) & flecha, porrete, arma de fogo, míssil, lança, machado, punhal, arpão, pistola, espada, lâmina, dinamite, machadinha, rifle, tanque, bomba, arma de fogo, faca, espingarda, gás lacrimogêneo, canhão, granada, maça, estilingue, chicote\\ \hline
 pleasant &  (attributes) & carícia, liberdade, saúde, amor, diploma, paz, prazer, alegrar, amigo, céu, leal,diamante, gentil, honesto, sortudo, arco-íris, prenda, honra, milagre, amanhecer, família, feliz,riso, paraíso, férias\\  \hline
unpleasant &  (attributes) & maus-tratos, colisão, imundíce, assassinato, enfermidade, acidente, morte, tristeza, veneno, fedor, assalto, desastre, ódio, tragédia, poluir, divórcio, cadeia, pobreza, feio, cancro, matar, divórcio, cadeia, pobreza, feio, cancro, matar, podre, vómito, agonia, prisão\\ 
 \hline
\end{tabular}
\caption{Portugese Stimuli (Word List)}
\label{tab:portugese}
\end{table*}
\end{center}


\begin{center}
\begin{table*}
\begin{tabular}{ |r l |p{0.7\textwidth}| } 
 \hline
  \multicolumn{2}{|c|}{\textbf{Category}}&\textbf{Spanish Stimuli}  \\ 
 \hline 
 flowers&  (target) & trébol, orquídea, rosa, narciso, lila, tulipán, margarita, lirio, violeta, magnolia  \\  \hline
 insects & (target) & hormiga, pulga, araña, ácaro, mosca, tarántula, abeja, cucaracha, mosquito, avispón \\  \hline
 instruments &  (target) & cornamusa, violonchelo, guitarra, flauta, trombón, banjo, clarinete, harmónica, mandolina, trompeta, fagot, tambor, arpa, oboe, tuba, campana, fiddle, clave, piano, viola, bongo, flute, cuerno, saxofón, violín\\  \hline
 weapons &  (target) & flecha, palo, pistola, misil, lanza, hacha, daga, arpón, espada, cuchilla, dinamitar, rifle, tanque, bomba, naja, escopeta, cañón, granada, mazo, honda, látigo\\  \hline
 pleasant &  (attributes) & caricia, libertad, salud, amor, diploma, paz, placer, ánimo, amigo, cielo, leal, diamante, delicado, honesto, afortunado, arco-iris, obsequio, honor, milagro, amanecer, familia, feliz\\  \hline
unpleasant &  (attributes) & maltrato, choque, inmundicia, asesinato, enfermedad, accidente, muerte, pena, ponzoña, hedor, asalto, desastre, odio, contaminar, tragedia, divorcio, cárcel, pobreza, feo, cáncer, matar, podrido, vómito, agonía, prisión\\  \hline
 
\end{tabular}
\caption{Spanish Stimuli (Word List)}
\label{tab:spanish}
\end{table*}
\end{center}


\begin{center}
\begin{table*}
\begin{tabular}{ |r l|p{0.7\textwidth}| } 
 \hline
  \multicolumn{2}{|c|}{\textbf{Category}}&\textbf{Turkish Stimuli}  \\ 
 \hline 
 flowers&  (target) & yonca, orkide, gül, nergis, leylak, lale, papatya, zambak, menekşe, manolya  \\  \hline
 insects & (target) & karınca, pire, örümcek, tahtakurusu, sinek, tarantula, arı, hamamböceği, sivrisinek, eşekarısı  \\  \hline
 instruments &  (target) & gayda,çello, gitar, ut, trombon, banço, klarnet, mızıka, mandolin, trompet, fagot, davul, arp, obua, tuba, zil, keman, harpsikord, piyano, viyola, tamtam, flüt, boynuz, saksafon, viyolin\\ \hline
 weapons &  (target) & ok, cop, tabanca, mermi, mızrak, balta, hançer, zıpkın, silah, kılıç, bıçak, dinamit, nacak, tüfek, tank, bomba, silâh, bıçak, çifte, gözyaşı gazı, gülle, bombası, topuz, mancınık, kırbaç\\ \hline
 pleasant &  (attributes) & okşamak, özgürlük, sağlık, sevgi, barış, neşe, arkadaş,cennet, sadık, keyif, pırlanta, kibar, dürüst, şanslı, gökkuşağı,diploma, hediye, onur, mucize, gündoğumu, aile, mutlu, kahkaha,cennet, tatil\\ \hline
unpleasant &  (attributes) & istismar, çarpmak  pislik  cinayet, hastalık, ölüm , üzüntü , zehir , kokuşmuş , saldırı , felaket , nefret , kirletmek , facia , boşanmak , hapishane , fakirlik , çirkin , kanser , öldürmek , çürümüş , kusmuk , ızdırap , sancı, cezaevi\\ 
 \hline
\end{tabular}
\caption{Turkish Stimuli (Word List)}
\label{tab:turkish}
\end{table*}
\end{center}

\end{appendix}

%% file: main.bbl

\begin{thebibliography}{57}


\ifx \showCODEN    \undefined \def \showCODEN     #1{\unskip}     \fi
\ifx \showDOI      \undefined \def \showDOI       #1{#1}\fi
\ifx \showISBNx    \undefined \def \showISBNx     #1{\unskip}     \fi
\ifx \showISBNxiii \undefined \def \showISBNxiii  #1{\unskip}     \fi
\ifx \showISSN     \undefined \def \showISSN      #1{\unskip}     \fi
\ifx \showLCCN     \undefined \def \showLCCN      #1{\unskip}     \fi
\ifx \shownote     \undefined \def \shownote      #1{#1}          \fi
\ifx \showarticletitle \undefined \def \showarticletitle #1{#1}   \fi
\ifx \showURL      \undefined \def \showURL       {\relax}        \fi
\providecommand\bibfield[2]{#2}
\providecommand\bibinfo[2]{#2}
\providecommand\natexlab[1]{#1}
\providecommand\showeprint[2][]{arXiv:#2}

\bibitem[\protect\citeauthoryear{Bakarov}{Bakarov}{2018}]%
        {Bakarov}
\bibfield{author}{\bibinfo{person}{Amir Bakarov}.}
  \bibinfo{year}{2018}\natexlab{}.
\newblock \showarticletitle{A survey of word embeddings evaluation methods}.
\newblock \bibinfo{journal}{\emph{arXiv preprint arXiv:1801.09536}}
  (\bibinfo{year}{2018}).
\newblock


\bibitem[\protect\citeauthoryear{Bellezza, Greenwald, and Banaji}{Bellezza
  et~al\mbox{.}}{1986}]%
        {bellezza1986words}
\bibfield{author}{\bibinfo{person}{Francis~S Bellezza},
  \bibinfo{person}{Anthony~G Greenwald}, {and} \bibinfo{person}{Mahzarin~R
  Banaji}.} \bibinfo{year}{1986}\natexlab{}.
\newblock \showarticletitle{Words high and low in pleasantness as rated by male
  and female college students}.
\newblock \bibinfo{journal}{\emph{Behavior Research Methods, Instruments, \&
  Computers}} \bibinfo{volume}{18}, \bibinfo{number}{3} (\bibinfo{year}{1986}),
  \bibinfo{pages}{299--303}.
\newblock


\bibitem[\protect\citeauthoryear{Bojanowski, Grave, Joulin, and
  Mikolov}{Bojanowski et~al\mbox{.}}{2016}]%
        {Bojanowski}
\bibfield{author}{\bibinfo{person}{Piotr Bojanowski}, \bibinfo{person}{Edouard
  Grave}, \bibinfo{person}{Armand Joulin}, {and} \bibinfo{person}{Tomas
  Mikolov}.} \bibinfo{year}{2016}\natexlab{}.
\newblock \showarticletitle{Enriching Word Vectors with Subword Information}.
\newblock \bibinfo{journal}{\emph{Transactions of the Association for
  Computational Linguistics}}  \bibinfo{volume}{5} (\bibinfo{date}{07}
  \bibinfo{year}{2016}).
\newblock


\bibitem[\protect\citeauthoryear{Bradley and Lang}{Bradley and Lang}{1999}]%
        {bradley1999affective}
\bibfield{author}{\bibinfo{person}{Margaret~M Bradley} {and}
  \bibinfo{person}{Peter~J Lang}.} \bibinfo{year}{1999}\natexlab{}.
\newblock \bibinfo{booktitle}{\emph{Affective norms for English words (ANEW):
  Instruction manual and affective ratings}}.
\newblock \bibinfo{type}{{T}echnical {R}eport}. \bibinfo{institution}{Technical
  report C-1, the center for research in psychophysiology, University of
  Florida}.
\newblock


\bibitem[\protect\citeauthoryear{Bruni, Tran, and Baroni}{Bruni
  et~al\mbox{.}}{2014}]%
        {bruni2014multimodal}
\bibfield{author}{\bibinfo{person}{Elia Bruni}, \bibinfo{person}{Nam-Khanh
  Tran}, {and} \bibinfo{person}{Marco Baroni}.}
  \bibinfo{year}{2014}\natexlab{}.
\newblock \showarticletitle{Multimodal distributional semantics}.
\newblock \bibinfo{journal}{\emph{Journal of Artificial Intelligence Research}}
   \bibinfo{volume}{49} (\bibinfo{year}{2014}), \bibinfo{pages}{1--47}.
\newblock


\bibitem[\protect\citeauthoryear{Caliskan, Bryson, and Narayanan}{Caliskan
  et~al\mbox{.}}{2017}]%
        {caliskan}
\bibfield{author}{\bibinfo{person}{Aylin Caliskan}, \bibinfo{person}{Joanna
  Bryson}, {and} \bibinfo{person}{Arvind Narayanan}.}
  \bibinfo{year}{2017}\natexlab{}.
\newblock \showarticletitle{Semantics derived automatically from language
  corpora contain human-like biases}.
\newblock \bibinfo{journal}{\emph{Science}}  \bibinfo{volume}{356}
  (\bibinfo{date}{04} \bibinfo{year}{2017}), \bibinfo{pages}{183--186}.
\newblock


\bibitem[\protect\citeauthoryear{Caliskan and Lewis}{Caliskan and
  Lewis}{2020}]%
        {caliskan2020social}
\bibfield{author}{\bibinfo{person}{Aylin Caliskan} {and} \bibinfo{person}{Molly
  Lewis}.} \bibinfo{year}{2020}\natexlab{}.
\newblock \showarticletitle{Social biases in word embeddings and their relation
  to human cognition}.
\newblock  (\bibinfo{year}{2020}).
\newblock


\bibitem[\protect\citeauthoryear{Cohen}{Cohen}{2013}]%
        {cohen2013statistical}
\bibfield{author}{\bibinfo{person}{Jacob Cohen}.}
  \bibinfo{year}{2013}\natexlab{}.
\newblock \bibinfo{booktitle}{\emph{Statistical power analysis for the
  behavioral sciences}}.
\newblock \bibinfo{publisher}{Academic press}.
\newblock


\bibitem[\protect\citeauthoryear{Dodge}{Dodge}{2008}]%
        {spearman}
\bibfield{author}{\bibinfo{person}{Yadolah Dodge}.}
  \bibinfo{year}{2008}\natexlab{}.
\newblock \showarticletitle{Spearman rank correlation coefficient}.
\newblock \bibinfo{journal}{\emph{The Concise Encyclopedia of Statistics.
  Springer, New York}} (\bibinfo{year}{2008}), \bibinfo{pages}{502--505}.
\newblock


\bibitem[\protect\citeauthoryear{Ethayarajh, Duvenaud, and Hirst}{Ethayarajh
  et~al\mbox{.}}{2019}]%
        {ethayarajh2019understanding}
\bibfield{author}{\bibinfo{person}{Kawin Ethayarajh}, \bibinfo{person}{David
  Duvenaud}, {and} \bibinfo{person}{Graeme Hirst}.}
  \bibinfo{year}{2019}\natexlab{}.
\newblock \showarticletitle{Understanding undesirable word embedding
  associations}.
\newblock \bibinfo{journal}{\emph{arXiv preprint arXiv:1908.06361}}
  (\bibinfo{year}{2019}).
\newblock


\bibitem[\protect\citeauthoryear{Faruqui, Tsvetkov, Rastogi, and Dyer}{Faruqui
  et~al\mbox{.}}{2016}]%
        {faruqui-etal-2016-problems}
\bibfield{author}{\bibinfo{person}{Manaal Faruqui}, \bibinfo{person}{Yulia
  Tsvetkov}, \bibinfo{person}{Pushpendre Rastogi}, {and} \bibinfo{person}{Chris
  Dyer}.} \bibinfo{year}{2016}\natexlab{}.
\newblock \showarticletitle{Problems With Evaluation of Word Embeddings Using
  Word Similarity Tasks}. In \bibinfo{booktitle}{\emph{Proceedings of the 1st
  Workshop on Evaluating Vector-Space Representations for {NLP}}}.
  \bibinfo{publisher}{Association for Computational Linguistics},
  \bibinfo{address}{Berlin, Germany}, \bibinfo{pages}{30--35}.
\newblock
\urldef\tempurl%
\url{https://www.aclweb.org/anthology/W16-2506}
\showURL{%
\tempurl}


\bibitem[\protect\citeauthoryear{Finkelstein, Gabrilovich, Matias, Rivlin,
  Solan, Wolfman, and Ruppin}{Finkelstein et~al\mbox{.}}{2001}]%
        {finkelstein2001placing}
\bibfield{author}{\bibinfo{person}{Lev Finkelstein}, \bibinfo{person}{Evgeniy
  Gabrilovich}, \bibinfo{person}{Yossi Matias}, \bibinfo{person}{Ehud Rivlin},
  \bibinfo{person}{Zach Solan}, \bibinfo{person}{Gadi Wolfman}, {and}
  \bibinfo{person}{Eytan Ruppin}.} \bibinfo{year}{2001}\natexlab{}.
\newblock \showarticletitle{Placing search in context: The concept revisited}.
  In \bibinfo{booktitle}{\emph{Proceedings of the 10th international conference
  on World Wide Web}}. \bibinfo{pages}{406--414}.
\newblock


\bibitem[\protect\citeauthoryear{Frijda et~al\mbox{.}}{Frijda
  et~al\mbox{.}}{1986}]%
        {frijda1986emotions}
\bibfield{author}{\bibinfo{person}{Nico~H Frijda} {et~al\mbox{.}}}
  \bibinfo{year}{1986}\natexlab{}.
\newblock \bibinfo{booktitle}{\emph{The emotions}}.
\newblock \bibinfo{publisher}{Cambridge University Press}.
\newblock


\bibitem[\protect\citeauthoryear{Garg, Schiebinger, Jurafsky, and Zou}{Garg
  et~al\mbox{.}}{2018}]%
        {garg2018word}
\bibfield{author}{\bibinfo{person}{Nikhil Garg}, \bibinfo{person}{Londa
  Schiebinger}, \bibinfo{person}{Dan Jurafsky}, {and} \bibinfo{person}{James
  Zou}.} \bibinfo{year}{2018}\natexlab{}.
\newblock \showarticletitle{Word embeddings quantify 100 years of gender and
  ethnic stereotypes}.
\newblock \bibinfo{journal}{\emph{Proceedings of the National Academy of
  Sciences}} \bibinfo{volume}{115}, \bibinfo{number}{16}
  (\bibinfo{year}{2018}), \bibinfo{pages}{E3635--E3644}.
\newblock


\bibitem[\protect\citeauthoryear{Gerz, Vuli{\'c}, Hill, Reichart, and
  Korhonen}{Gerz et~al\mbox{.}}{2016}]%
        {gerz2016simverb}
\bibfield{author}{\bibinfo{person}{Daniela Gerz}, \bibinfo{person}{Ivan
  Vuli{\'c}}, \bibinfo{person}{Felix Hill}, \bibinfo{person}{Roi Reichart},
  {and} \bibinfo{person}{Anna Korhonen}.} \bibinfo{year}{2016}\natexlab{}.
\newblock \showarticletitle{Simverb-3500: A large-scale evaluation set of verb
  similarity}.
\newblock \bibinfo{journal}{\emph{arXiv preprint arXiv:1608.00869}}
  (\bibinfo{year}{2016}).
\newblock


\bibitem[\protect\citeauthoryear{Grave, Bojanowski, Gupta, Joulin, and
  Mikolov}{Grave et~al\mbox{.}}{2018}]%
        {grave2018learning}
\bibfield{author}{\bibinfo{person}{Edouard Grave}, \bibinfo{person}{Piotr
  Bojanowski}, \bibinfo{person}{Prakhar Gupta}, \bibinfo{person}{Armand
  Joulin}, {and} \bibinfo{person}{Tomas Mikolov}.}
  \bibinfo{year}{2018}\natexlab{}.
\newblock \showarticletitle{Learning word vectors for 157 languages}.
\newblock \bibinfo{journal}{\emph{arXiv preprint arXiv:1802.06893}}
  (\bibinfo{year}{2018}).
\newblock


\bibitem[\protect\citeauthoryear{Greenwald, McGhee, and Schwartz}{Greenwald
  et~al\mbox{.}}{1998}]%
        {Greenwald1998MeasuringID}
\bibfield{author}{\bibinfo{person}{Anthony~G Greenwald},
  \bibinfo{person}{Debbie~E. McGhee}, {and} \bibinfo{person}{Joe~L. Schwartz}.}
  \bibinfo{year}{1998}\natexlab{}.
\newblock \showarticletitle{Measuring individual differences in implicit
  cognition: the implicit association test.}
\newblock \bibinfo{journal}{\emph{Journal of personality and social
  psychology}}  \bibinfo{volume}{74 6} (\bibinfo{year}{1998}),
  \bibinfo{pages}{1464--80}.
\newblock


\bibitem[\protect\citeauthoryear{Guo and Caliskan}{Guo and Caliskan}{2021}]%
        {guo2021detecting}
\bibfield{author}{\bibinfo{person}{Wei Guo} {and} \bibinfo{person}{Aylin
  Caliskan}.} \bibinfo{year}{2021}\natexlab{}.
\newblock \showarticletitle{Detecting emergent intersectional biases:
  Contextualized word embeddings contain a distribution of human-like biases}.
  In \bibinfo{booktitle}{\emph{Proceedings of the 2021 AAAI/ACM Conference on
  AI, Ethics, and Society}}. \bibinfo{pages}{122--133}.
\newblock


\bibitem[\protect\citeauthoryear{Hamilton, Leskovec, and Jurafsky}{Hamilton
  et~al\mbox{.}}{2016}]%
        {hamilton2016diachronic}
\bibfield{author}{\bibinfo{person}{William~L Hamilton}, \bibinfo{person}{Jure
  Leskovec}, {and} \bibinfo{person}{Dan Jurafsky}.}
  \bibinfo{year}{2016}\natexlab{}.
\newblock \showarticletitle{Diachronic word embeddings reveal statistical laws
  of semantic change}.
\newblock \bibinfo{journal}{\emph{arXiv preprint arXiv:1605.09096}}
  (\bibinfo{year}{2016}).
\newblock


\bibitem[\protect\citeauthoryear{Harmon-Jones, Gable, and Price}{Harmon-Jones
  et~al\mbox{.}}{2013}]%
        {harmon2013does}
\bibfield{author}{\bibinfo{person}{Eddie Harmon-Jones},
  \bibinfo{person}{Philip~A Gable}, {and} \bibinfo{person}{Tom~F Price}.}
  \bibinfo{year}{2013}\natexlab{}.
\newblock \showarticletitle{Does negative affect always narrow and positive
  affect always broaden the mind? Considering the influence of motivational
  intensity on cognitive scope}.
\newblock \bibinfo{journal}{\emph{Current Directions in Psychological Science}}
  \bibinfo{volume}{22}, \bibinfo{number}{4} (\bibinfo{year}{2013}),
  \bibinfo{pages}{301--307}.
\newblock


\bibitem[\protect\citeauthoryear{Hatzivassiloglou and McKeown}{Hatzivassiloglou
  and McKeown}{1997}]%
        {hatzivassiloglou1997predicting}
\bibfield{author}{\bibinfo{person}{Vasileios Hatzivassiloglou} {and}
  \bibinfo{person}{Kathleen McKeown}.} \bibinfo{year}{1997}\natexlab{}.
\newblock \showarticletitle{Predicting the semantic orientation of adjectives}.
  In \bibinfo{booktitle}{\emph{35th annual meeting of the association for
  computational linguistics and 8th conference of the european chapter of the
  association for computational linguistics}}. \bibinfo{pages}{174--181}.
\newblock


\bibitem[\protect\citeauthoryear{Hill, Reichart, and Korhonen}{Hill
  et~al\mbox{.}}{2015}]%
        {hill2015simlex}
\bibfield{author}{\bibinfo{person}{Felix Hill}, \bibinfo{person}{Roi Reichart},
  {and} \bibinfo{person}{Anna Korhonen}.} \bibinfo{year}{2015}\natexlab{}.
\newblock \showarticletitle{Simlex-999: Evaluating semantic models with
  (genuine) similarity estimation}.
\newblock \bibinfo{journal}{\emph{Computational Linguistics}}
  \bibinfo{volume}{41}, \bibinfo{number}{4} (\bibinfo{year}{2015}),
  \bibinfo{pages}{665--695}.
\newblock


\bibitem[\protect\citeauthoryear{Hollenstein, de~la Torre, Langer, and
  Zhang}{Hollenstein et~al\mbox{.}}{2019}]%
        {Hollenstein}
\bibfield{author}{\bibinfo{person}{Nora Hollenstein}, \bibinfo{person}{Antonio
  de~la Torre}, \bibinfo{person}{Nicolas Langer}, {and} \bibinfo{person}{Ce
  Zhang}.} \bibinfo{year}{2019}\natexlab{}.
\newblock \showarticletitle{CogniVal: A Framework for Cognitive Word Embedding
  Evaluation}. In \bibinfo{booktitle}{\emph{Proceedings of the 23rd Conference
  on Computational Natural Language Learning}}. \bibinfo{pages}{538--549}.
\newblock


\bibitem[\protect\citeauthoryear{Imbir}{Imbir}{2015}]%
        {imbir2015affective}
\bibfield{author}{\bibinfo{person}{Kamil~K Imbir}.}
  \bibinfo{year}{2015}\natexlab{}.
\newblock \showarticletitle{Affective norms for 1,586 polish words (ANPW):
  Duality-of-mind approach}.
\newblock \bibinfo{journal}{\emph{Behavior research methods}}
  \bibinfo{volume}{47}, \bibinfo{number}{3} (\bibinfo{year}{2015}),
  \bibinfo{pages}{860--870}.
\newblock


\bibitem[\protect\citeauthoryear{Kapucu, K{\i}l{\i}{\c{c}},
  {\"O}zk{\i}l{\i}{\c{c}}, and Sar{\i}baz}{Kapucu et~al\mbox{.}}{2018}]%
        {kapucu2018turkish}
\bibfield{author}{\bibinfo{person}{Aycan Kapucu}, \bibinfo{person}{Asl{\i}
  K{\i}l{\i}{\c{c}}}, \bibinfo{person}{Y{\i}ld{\i}z {\"O}zk{\i}l{\i}{\c{c}}},
  {and} \bibinfo{person}{Bengisu Sar{\i}baz}.} \bibinfo{year}{2018}\natexlab{}.
\newblock \showarticletitle{Turkish emotional word norms for arousal, valence,
  and discrete emotion categories}.
\newblock \bibinfo{journal}{\emph{Psychological reports}}
  (\bibinfo{year}{2018}), \bibinfo{pages}{0033294118814722}.
\newblock


\bibitem[\protect\citeauthoryear{Karpinski and Steinman}{Karpinski and
  Steinman}{2006}]%
        {karpinski2006single}
\bibfield{author}{\bibinfo{person}{Andrew Karpinski} {and}
  \bibinfo{person}{Ross~B Steinman}.} \bibinfo{year}{2006}\natexlab{}.
\newblock \showarticletitle{The single category implicit association test as a
  measure of implicit social cognition.}
\newblock \bibinfo{journal}{\emph{Journal of personality and social
  psychology}} \bibinfo{volume}{91}, \bibinfo{number}{1}
  (\bibinfo{year}{2006}), \bibinfo{pages}{16}.
\newblock


\bibitem[\protect\citeauthoryear{Kirch}{Kirch}{2008}]%
        {pearson}
\bibfield{editor}{\bibinfo{person}{Wilhelm Kirch}} (Ed.).
  \bibinfo{year}{2008}\natexlab{}.
\newblock \bibinfo{booktitle}{\emph{Pearson's Correlation Coefficient}}.
\newblock \bibinfo{publisher}{Springer Netherlands},
  \bibinfo{address}{Dordrecht}, \bibinfo{pages}{1090--1091}.
\newblock
\showISBNx{978-1-4020-5614-7}
\urldef\tempurl%
\url{https://doi.org/10.1007/978-1-4020-5614-7_2569}
\showDOI{\tempurl}


\bibitem[\protect\citeauthoryear{Lewis and Lupyan}{Lewis and Lupyan}{2020}]%
        {lewis2020gender}
\bibfield{author}{\bibinfo{person}{Molly Lewis} {and} \bibinfo{person}{Gary
  Lupyan}.} \bibinfo{year}{2020}\natexlab{}.
\newblock \showarticletitle{Gender stereotypes are reflected in the
  distributional structure of 25 languages}.
\newblock \bibinfo{journal}{\emph{Nature human behaviour}}
  (\bibinfo{year}{2020}), \bibinfo{pages}{1--8}.
\newblock


\bibitem[\protect\citeauthoryear{Li, Lu, Long, and Gui}{Li
  et~al\mbox{.}}{2017}]%
        {li}
\bibfield{author}{\bibinfo{person}{Minglei Li}, \bibinfo{person}{Qin Lu},
  \bibinfo{person}{Yunfei Long}, {and} \bibinfo{person}{Lin Gui}.}
  \bibinfo{year}{2017}\natexlab{}.
\newblock \showarticletitle{Inferring Affective Meanings of Words from Word
  Embedding}.
\newblock \bibinfo{journal}{\emph{IEEE Transactions on Affective Computing}}
  \bibinfo{volume}{PP} (\bibinfo{date}{07} \bibinfo{year}{2017}),
  \bibinfo{pages}{1--1}.
\newblock
\urldef\tempurl%
\url{https://doi.org/10.1109/TAFFC.2017.2723012}
\showDOI{\tempurl}


\bibitem[\protect\citeauthoryear{Luong, Socher, and Manning}{Luong
  et~al\mbox{.}}{2013}]%
        {luong2013better}
\bibfield{author}{\bibinfo{person}{Minh-Thang Luong}, \bibinfo{person}{Richard
  Socher}, {and} \bibinfo{person}{Christopher~D Manning}.}
  \bibinfo{year}{2013}\natexlab{}.
\newblock \showarticletitle{Better word representations with recursive neural
  networks for morphology}. In \bibinfo{booktitle}{\emph{Proceedings of the
  Seventeenth Conference on Computational Natural Language Learning}}.
  \bibinfo{pages}{104--113}.
\newblock


\bibitem[\protect\citeauthoryear{Mikolov, Chen, Corrado, and Dean}{Mikolov
  et~al\mbox{.}}{2013}]%
        {Mikolov2013EfficientEO}
\bibfield{author}{\bibinfo{person}{Tomas Mikolov}, \bibinfo{person}{Kai Chen},
  \bibinfo{person}{Gregory~S. Corrado}, {and} \bibinfo{person}{Jeffrey Dean}.}
  \bibinfo{year}{2013}\natexlab{}.
\newblock \showarticletitle{Efficient Estimation of Word Representations in
  Vector Space}.
\newblock \bibinfo{journal}{\emph{CoRR}}  \bibinfo{volume}{abs/1301.3781}
  (\bibinfo{year}{2013}).
\newblock


\bibitem[\protect\citeauthoryear{Mohammad}{Mohammad}{2016}]%
        {mohammad2016sentiment}
\bibfield{author}{\bibinfo{person}{Saif~M Mohammad}.}
  \bibinfo{year}{2016}\natexlab{}.
\newblock \showarticletitle{Sentiment analysis: Detecting valence, emotions,
  and other affectual states from text}.
\newblock In \bibinfo{booktitle}{\emph{Emotion measurement}}.
  \bibinfo{publisher}{Elsevier}, \bibinfo{pages}{201--237}.
\newblock


\bibitem[\protect\citeauthoryear{Nosek, Banaji, and Greenwald}{Nosek
  et~al\mbox{.}}{2002}]%
        {nosek2002harvesting}
\bibfield{author}{\bibinfo{person}{Brian~A Nosek}, \bibinfo{person}{Mahzarin~R
  Banaji}, {and} \bibinfo{person}{Anthony~G Greenwald}.}
  \bibinfo{year}{2002}\natexlab{}.
\newblock \showarticletitle{Harvesting implicit group attitudes and beliefs
  from a demonstration web site.}
\newblock \bibinfo{journal}{\emph{Group Dynamics: Theory, Research, and
  Practice}} \bibinfo{volume}{6}, \bibinfo{number}{1} (\bibinfo{year}{2002}),
  \bibinfo{pages}{101}.
\newblock


\bibitem[\protect\citeauthoryear{Nosek, Smyth, Sriram, Lindner, Devos, Ayala,
  Bar-Anan, Bergh, Cai, Gonsalkorale, et~al\mbox{.}}{Nosek
  et~al\mbox{.}}{2009}]%
        {nosek2009national}
\bibfield{author}{\bibinfo{person}{Brian~A Nosek}, \bibinfo{person}{Frederick~L
  Smyth}, \bibinfo{person}{Natarajan Sriram}, \bibinfo{person}{Nicole~M
  Lindner}, \bibinfo{person}{Thierry Devos}, \bibinfo{person}{Alfonso Ayala},
  \bibinfo{person}{Yoav Bar-Anan}, \bibinfo{person}{Robin Bergh},
  \bibinfo{person}{Huajian Cai}, \bibinfo{person}{Karen Gonsalkorale},
  {et~al\mbox{.}}} \bibinfo{year}{2009}\natexlab{}.
\newblock \showarticletitle{National differences in gender--science stereotypes
  predict national sex differences in science and math achievement}.
\newblock \bibinfo{journal}{\emph{Proceedings of the National Academy of
  Sciences}} \bibinfo{volume}{106}, \bibinfo{number}{26}
  (\bibinfo{year}{2009}), \bibinfo{pages}{10593--10597}.
\newblock


\bibitem[\protect\citeauthoryear{Osgood}{Osgood}{1964}]%
        {osgood1964semantic}
\bibfield{author}{\bibinfo{person}{Charles~E Osgood}.}
  \bibinfo{year}{1964}\natexlab{}.
\newblock \showarticletitle{Semantic differential technique in the comparative
  study of cultures}.
\newblock \bibinfo{journal}{\emph{American Anthropologist}}
  \bibinfo{volume}{66}, \bibinfo{number}{3} (\bibinfo{year}{1964}),
  \bibinfo{pages}{171--200}.
\newblock


\bibitem[\protect\citeauthoryear{Osgood, May, Miron, and Miron}{Osgood
  et~al\mbox{.}}{1975}]%
        {osgood1975cross}
\bibfield{author}{\bibinfo{person}{Charles~Egerton Osgood},
  \bibinfo{person}{William~H May}, \bibinfo{person}{Murray~Samuel Miron}, {and}
  \bibinfo{person}{Murray~S Miron}.} \bibinfo{year}{1975}\natexlab{}.
\newblock \bibinfo{booktitle}{\emph{Cross-cultural universals of affective
  meaning}}. Vol.~\bibinfo{volume}{1}.
\newblock \bibinfo{publisher}{University of Illinois Press}.
\newblock


\bibitem[\protect\citeauthoryear{Osgood, Suci, and Tannenbaum}{Osgood
  et~al\mbox{.}}{1957}]%
        {osgood1957measurement}
\bibfield{author}{\bibinfo{person}{Charles~Egerton Osgood},
  \bibinfo{person}{George~J Suci}, {and} \bibinfo{person}{Percy~H Tannenbaum}.}
  \bibinfo{year}{1957}\natexlab{}.
\newblock \bibinfo{booktitle}{\emph{The measurement of meaning}}.
\newblock Number~47. \bibinfo{publisher}{University of Illinois press}.
\newblock


\bibitem[\protect\citeauthoryear{Paridon and Thompson}{Paridon and
  Thompson}{2019}]%
        {Paridon}
\bibfield{author}{\bibinfo{person}{Jeroen Paridon} {and} \bibinfo{person}{Bill
  Thompson}.} \bibinfo{year}{2019}\natexlab{}.
\newblock \bibinfo{title}{subs2vec: Word embeddings from subtitles in 55
  languages}.
\newblock
\newblock


\bibitem[\protect\citeauthoryear{Pennington, Socher, and Manning}{Pennington
  et~al\mbox{.}}{2014}]%
        {pennington}
\bibfield{author}{\bibinfo{person}{Jeffrey Pennington},
  \bibinfo{person}{Richard Socher}, {and} \bibinfo{person}{Christopher~D
  Manning}.} \bibinfo{year}{2014}\natexlab{}.
\newblock \showarticletitle{Glove: Global vectors for word representation}. In
  \bibinfo{booktitle}{\emph{Proceedings of the 2014 conference on empirical
  methods in natural language processing (EMNLP)}}.
  \bibinfo{pages}{1532--1543}.
\newblock


\bibitem[\protect\citeauthoryear{Redondo, Fraga, Padr{\'o}n, and
  Comesa{\~n}a}{Redondo et~al\mbox{.}}{2007}]%
        {redondo2007spanish}
\bibfield{author}{\bibinfo{person}{Jaime Redondo}, \bibinfo{person}{Isabel
  Fraga}, \bibinfo{person}{Isabel Padr{\'o}n}, {and}
  \bibinfo{person}{Montserrat Comesa{\~n}a}.} \bibinfo{year}{2007}\natexlab{}.
\newblock \showarticletitle{The Spanish adaptation of ANEW (affective norms for
  English words)}.
\newblock \bibinfo{journal}{\emph{Behavior research methods}}
  \bibinfo{volume}{39}, \bibinfo{number}{3} (\bibinfo{year}{2007}),
  \bibinfo{pages}{600--605}.
\newblock


\bibitem[\protect\citeauthoryear{Riloff and Wiebe}{Riloff and Wiebe}{2003}]%
        {riloff2003learning}
\bibfield{author}{\bibinfo{person}{Ellen Riloff} {and} \bibinfo{person}{Janyce
  Wiebe}.} \bibinfo{year}{2003}\natexlab{}.
\newblock \showarticletitle{Learning extraction patterns for subjective
  expressions}. In \bibinfo{booktitle}{\emph{Proceedings of the 2003 conference
  on Empirical methods in natural language processing}}.
  \bibinfo{pages}{105--112}.
\newblock


\bibitem[\protect\citeauthoryear{Russell and Mehrabian}{Russell and
  Mehrabian}{1977}]%
        {russell}
\bibfield{author}{\bibinfo{person}{James Russell} {and} \bibinfo{person}{Albert
  Mehrabian}.} \bibinfo{year}{1977}\natexlab{}.
\newblock \showarticletitle{Evidence for a Three-Factor Theory of Emotions}.
\newblock \bibinfo{journal}{\emph{Journal of Research in Personality}}
  \bibinfo{volume}{11} (\bibinfo{date}{09} \bibinfo{year}{1977}),
  \bibinfo{pages}{273--294}.
\newblock
\urldef\tempurl%
\url{https://doi.org/10.1016/0092-6566(77)90037-X}
\showDOI{\tempurl}


\bibitem[\protect\citeauthoryear{Russell}{Russell}{1983}]%
        {russell1983pancultural}
\bibfield{author}{\bibinfo{person}{James~A Russell}.}
  \bibinfo{year}{1983}\natexlab{}.
\newblock \showarticletitle{Pancultural aspects of the human conceptual
  organization of emotions.}
\newblock \bibinfo{journal}{\emph{Journal of personality and social
  psychology}} \bibinfo{volume}{45}, \bibinfo{number}{6}
  (\bibinfo{year}{1983}), \bibinfo{pages}{1281}.
\newblock


\bibitem[\protect\citeauthoryear{Schmidtke, Schr{\"o}der, Jacobs, and
  Conrad}{Schmidtke et~al\mbox{.}}{2014}]%
        {schmidtke2014angst}
\bibfield{author}{\bibinfo{person}{David~S Schmidtke}, \bibinfo{person}{Tobias
  Schr{\"o}der}, \bibinfo{person}{Arthur~M Jacobs}, {and}
  \bibinfo{person}{Markus Conrad}.} \bibinfo{year}{2014}\natexlab{}.
\newblock \showarticletitle{ANGST: Affective norms for German sentiment terms,
  derived from the affective norms for English words}.
\newblock \bibinfo{journal}{\emph{Behavior research methods}}
  \bibinfo{volume}{46}, \bibinfo{number}{4} (\bibinfo{year}{2014}),
  \bibinfo{pages}{1108--1118}.
\newblock


\bibitem[\protect\citeauthoryear{Schnabel, Labutov, Mimno, and
  Joachims}{Schnabel et~al\mbox{.}}{2015}]%
        {Schnabel2015}
\bibfield{author}{\bibinfo{person}{Tobias Schnabel}, \bibinfo{person}{Igor
  Labutov}, \bibinfo{person}{David~M. Mimno}, {and} \bibinfo{person}{Thorsten
  Joachims}.} \bibinfo{year}{2015}\natexlab{}.
\newblock \showarticletitle{Evaluation methods for unsupervised word
  embeddings}. In \bibinfo{booktitle}{\emph{EMNLP}}.
\newblock


\bibitem[\protect\citeauthoryear{Soares, Comesa{\~n}a, Pinheiro, Sim{\~o}es,
  and Frade}{Soares et~al\mbox{.}}{2012}]%
        {soares2012adaptation}
\bibfield{author}{\bibinfo{person}{Ana~Paula Soares},
  \bibinfo{person}{Montserrat Comesa{\~n}a}, \bibinfo{person}{Ana~P Pinheiro},
  \bibinfo{person}{Alberto Sim{\~o}es}, {and} \bibinfo{person}{Carla~Sofia
  Frade}.} \bibinfo{year}{2012}\natexlab{}.
\newblock \showarticletitle{The adaptation of the Affective Norms for English
  words (ANEW) for European Portuguese}.
\newblock \bibinfo{journal}{\emph{Behavior research methods}}
  \bibinfo{volume}{44}, \bibinfo{number}{1} (\bibinfo{year}{2012}),
  \bibinfo{pages}{256--269}.
\newblock


\bibitem[\protect\citeauthoryear{Speer, Chin, and Havasi}{Speer
  et~al\mbox{.}}{2017}]%
        {speer2017conceptnet}
\bibfield{author}{\bibinfo{person}{Robyn Speer}, \bibinfo{person}{Joshua Chin},
  {and} \bibinfo{person}{Catherine Havasi}.} \bibinfo{year}{2017}\natexlab{}.
\newblock \showarticletitle{Conceptnet 5.5: An open multilingual graph of
  general knowledge}. In \bibinfo{booktitle}{\emph{Thirty-First AAAI Conference
  on Artificial Intelligence}}.
\newblock


\bibitem[\protect\citeauthoryear{Teofili and Chhaya}{Teofili and
  Chhaya}{2019}]%
        {teofili2019affect}
\bibfield{author}{\bibinfo{person}{Tommaso Teofili} {and}
  \bibinfo{person}{Niyati Chhaya}.} \bibinfo{year}{2019}\natexlab{}.
\newblock \showarticletitle{Affect Enriched Word Embeddings for News
  Information Retrieval}.
\newblock \bibinfo{journal}{\emph{arXiv preprint arXiv:1909.01772}}
  (\bibinfo{year}{2019}).
\newblock


\bibitem[\protect\citeauthoryear{Tsvetkov, Faruqui, and Dyer}{Tsvetkov
  et~al\mbox{.}}{2016}]%
        {Tsvetkov}
\bibfield{author}{\bibinfo{person}{Yulia Tsvetkov}, \bibinfo{person}{Manaal
  Faruqui}, {and} \bibinfo{person}{Chris Dyer}.}
  \bibinfo{year}{2016}\natexlab{}.
\newblock \showarticletitle{Correlation-based Intrinsic Evaluation of Word
  Vector Representations}. In \bibinfo{booktitle}{\emph{Proceedings of the 1st
  Workshop on Evaluating Vector-Space Representations for NLP}}.
  \bibinfo{pages}{111--115}.
\newblock


\bibitem[\protect\citeauthoryear{Turney and Littman}{Turney and
  Littman}{2003}]%
        {turney2003measuring}
\bibfield{author}{\bibinfo{person}{Peter~D Turney} {and}
  \bibinfo{person}{Michael~L Littman}.} \bibinfo{year}{2003}\natexlab{}.
\newblock \showarticletitle{Measuring praise and criticism: Inference of
  semantic orientation from association}.
\newblock \bibinfo{journal}{\emph{ACM Transactions on Information Systems
  (TOIS)}} \bibinfo{volume}{21}, \bibinfo{number}{4} (\bibinfo{year}{2003}),
  \bibinfo{pages}{315--346}.
\newblock


\bibitem[\protect\citeauthoryear{Turney and Pantel}{Turney and Pantel}{2010}]%
        {turney2010frequency}
\bibfield{author}{\bibinfo{person}{Peter~D Turney} {and}
  \bibinfo{person}{Patrick Pantel}.} \bibinfo{year}{2010}\natexlab{}.
\newblock \showarticletitle{From frequency to meaning: Vector space models of
  semantics}.
\newblock \bibinfo{journal}{\emph{Journal of artificial intelligence research}}
   \bibinfo{volume}{37} (\bibinfo{year}{2010}), \bibinfo{pages}{141--188}.
\newblock


\bibitem[\protect\citeauthoryear{Ungar, Preotiuc-Pietro, and Sedoc}{Ungar
  et~al\mbox{.}}{2017}]%
        {Ungar2017PredictingEW}
\bibfield{author}{\bibinfo{person}{Lyle~H. Ungar}, \bibinfo{person}{Daniel
  Preotiuc-Pietro}, {and} \bibinfo{person}{Jo{\~a}o Sedoc}.}
  \bibinfo{year}{2017}\natexlab{}.
\newblock \showarticletitle{Predicting Emotional Word Ratings using
  Distributional Representations and Signed Clustering}. In
  \bibinfo{booktitle}{\emph{EACL}}.
\newblock


\bibitem[\protect\citeauthoryear{Vo, Conrad, Kuchinke, Urton, Hofmann, and
  Jacobs}{Vo et~al\mbox{.}}{2009}]%
        {vo2009berlin}
\bibfield{author}{\bibinfo{person}{Melissa~LH Vo}, \bibinfo{person}{Markus
  Conrad}, \bibinfo{person}{Lars Kuchinke}, \bibinfo{person}{Karolina Urton},
  \bibinfo{person}{Markus~J Hofmann}, {and} \bibinfo{person}{Arthur~M Jacobs}.}
  \bibinfo{year}{2009}\natexlab{}.
\newblock \showarticletitle{The Berlin affective word list reloaded (BAWL-R)}.
\newblock \bibinfo{journal}{\emph{Behavior research methods}}
  \bibinfo{volume}{41}, \bibinfo{number}{2} (\bibinfo{year}{2009}),
  \bibinfo{pages}{534--538}.
\newblock


\bibitem[\protect\citeauthoryear{Wang, Wang, Chen, Wang, and Kuo}{Wang
  et~al\mbox{.}}{2019}]%
        {Wang}
\bibfield{author}{\bibinfo{person}{Bin Wang}, \bibinfo{person}{Angela Wang},
  \bibinfo{person}{Fenxiao Chen}, \bibinfo{person}{Yuncheng Wang}, {and}
  \bibinfo{person}{C.-C.~Jay Kuo}.} \bibinfo{year}{2019}\natexlab{}.
\newblock \showarticletitle{Evaluating Word Embedding Models: Methods and
  Experimental Results}.
\newblock \bibinfo{journal}{\emph{ArXiv}} (\bibinfo{year}{2019}).
\newblock


\bibitem[\protect\citeauthoryear{Wang, Yu, Lai, and Zhang}{Wang
  et~al\mbox{.}}{2016}]%
        {wang2016}
\bibfield{author}{\bibinfo{person}{Jin Wang}, \bibinfo{person}{Liang-Chih Yu},
  \bibinfo{person}{K. Lai}, {and} \bibinfo{person}{Xuejie Zhang}.}
  \bibinfo{year}{2016}\natexlab{}.
\newblock \showarticletitle{Community-Based Weighted Graph Model for
  Valence-Arousal Prediction of Affective Words}.
\newblock \bibinfo{journal}{\emph{IEEE/ACM Transactions on Audio, Speech, and
  Language Processing}}  \bibinfo{volume}{24} (\bibinfo{date}{11}
  \bibinfo{year}{2016}), \bibinfo{pages}{1--1}.
\newblock
\urldef\tempurl%
\url{https://doi.org/10.1109/TASLP.2016.2594287}
\showDOI{\tempurl}


\bibitem[\protect\citeauthoryear{Warriner, Kuperman, and Brysbaert}{Warriner
  et~al\mbox{.}}{2013}]%
        {warriner}
\bibfield{author}{\bibinfo{person}{Amy Warriner}, \bibinfo{person}{Victor
  Kuperman}, {and} \bibinfo{person}{Marc Brysbaert}.}
  \bibinfo{year}{2013}\natexlab{}.
\newblock \showarticletitle{Norms of valence, arousal, and dominance for 13,915
  English lemmas}.
\newblock \bibinfo{journal}{\emph{Behavior research methods}}
  \bibinfo{volume}{45} (\bibinfo{date}{02} \bibinfo{year}{2013}).
\newblock
\urldef\tempurl%
\url{https://doi.org/10.3758/s13428-012-0314-x}
\showDOI{\tempurl}


\bibitem[\protect\citeauthoryear{Yu, Lee, Hao, Wang, He, Hu, Lai, and Zhang}{Yu
  et~al\mbox{.}}{2016}]%
        {yu-etal-2016-building}
\bibfield{author}{\bibinfo{person}{Liang-Chih Yu}, \bibinfo{person}{Lung-Hao
  Lee}, \bibinfo{person}{Shuai Hao}, \bibinfo{person}{Jin Wang},
  \bibinfo{person}{Yunchao He}, \bibinfo{person}{Jun Hu},
  \bibinfo{person}{K.~Robert Lai}, {and} \bibinfo{person}{Xuejie Zhang}.}
  \bibinfo{year}{2016}\natexlab{}.
\newblock \showarticletitle{Building {C}hinese Affective Resources in
  Valence-Arousal Dimensions}. In \bibinfo{booktitle}{\emph{Proceedings of the
  2016 Conference of the North {A}merican Chapter of the Association for
  Computational Linguistics: Human Language Technologies}}.
  \bibinfo{publisher}{Association for Computational Linguistics},
  \bibinfo{address}{San Diego, California}, \bibinfo{pages}{540--545}.
\newblock


\end{thebibliography}
